\documentclass[amssymb,aps,prd,two column,nofootinbib,superscriptaddress]{revtex4-2}
\usepackage{graphicx}
\usepackage{subcaption} 
\usepackage{mathtools}
\usepackage{slashed}
\usepackage{url}
\usepackage[normalem]{ulem}
\usepackage{enumitem}
\usepackage{mathrsfs}
\usepackage{float}
\usepackage[colorlinks=true, linkcolor=blue, citecolor=red, urlcolor=magenta]{hyperref}
\newcommand{\orcid}[1]{\href{https://orcid.org/#1}{\includegraphics[width=8pt]
		{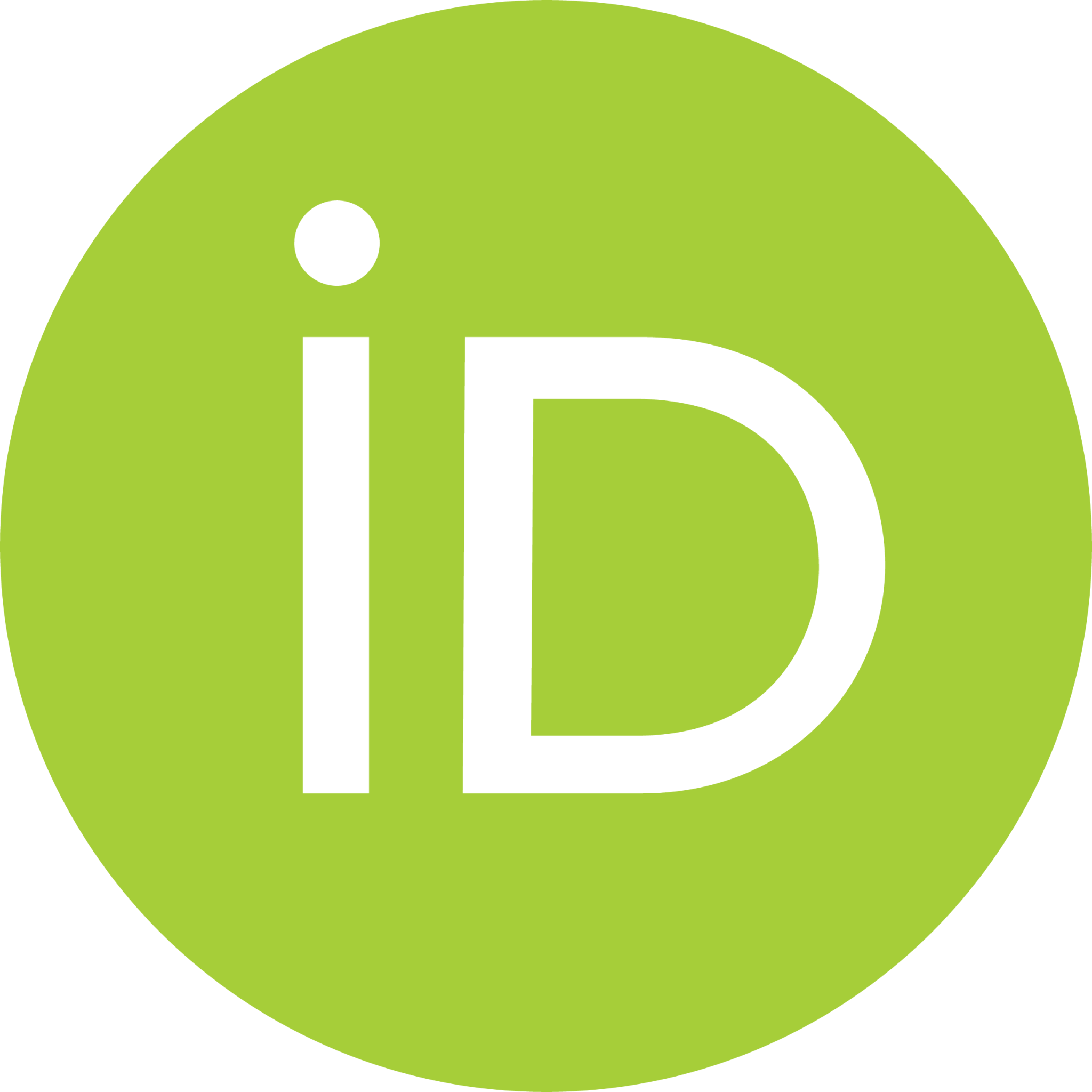}}}

\begin{document}
	
	\title{Towards compressed baryonic matter densities: $D$ meson diffusion}
	\author{Dani Rose J Marattukalam\orcid{0009-0006-8204-8148}}\email{danirosej@iitbhilai.ac.in}\affiliation{Department of Physics, Indian Institute of Technology Bhilai, Kutelabhata, Durg 491002, India} \author{Manpreet Kaur\orcid{0009-0003-7960-1625}}\email{ranapreeti803@gmail.com}\affiliation{Department of Physics, Dr. B R Ambedkar National Institute of Technology, Jalandhar – 144008, Punjab, India} \author{Arvind Kumar\orcid{0000-0003-1873-6094}}\email{kumara@nitj.ac.in}\affiliation{Department of Physics, Dr. B R Ambedkar National Institute of Technology, Jalandhar – 144008, Punjab, India} \author{Sabyasachi Ghosh\orcid{0000-0003-1212-824X}}\email{sabya@iitbhilai.ac.in}\affiliation{Department of Physics, Indian Institute of Technology Bhilai, Kutelabhata, Durg 491002, India}
	
\begin{abstract}
We study the spatial diffusion coefficient and the momentum transport coefficients of $D$ mesons through a dense nuclear medium in the relaxation time approximation of the kinetic theory. The in medium modifications of the $D$ meson transport properties are computed in the chiral SU(3) hadronic model. Relaxation time is estimated using dilute and degenerate gas approximations for low and high baryonic densities, respectively. We have noticed that relaxation time and spatial diffusion of $D$ meson decrease rapidly in the low density dilute gas domain and mildly in the high density degenerate gas domain. The detailed result of the present work on $D$ meson diffusion is quite contemporary and important towards the compressed baryonic matter densities which can be assessed in future heavy ion collision experiments. 

\end{abstract}
\maketitle

\section{Introduction}
The quark-gluon plasma (QGP) produced in relativistic heavy-ion collisions behaves as a nearly perfect fluid whose small viscosity rapidly thermalizes light quarks and gluons~\cite{Gossiaux:2009hr}. Heavy quarks, however, thermalize far more slowly, with non-equilibrated heavy quarks and mesons diffusing through the QGP and the subsequent hadronic medium~\cite{PhysRevC.82.014908}. As a result, heavy mesons retain information about the early QGP and its evolution, whereas near-thermal light hadrons mainly reflect conditions at the phase transition. Since heavy flavor is produced almost entirely in the initial stage with negligible thermal production, charm quarks and charmed mesons are reliable probes of the medium's transport properties, bridging early QGP dynamics and the hadronic final state~\cite{Rapp:2008qc,Prino:2016cni}. Heavy flavor further probes the full transverse-momentum range---Langevin diffusion at low $p_T$, fragmentation and coalescence at intermediate $p_T$, and radiative loss and jet quenching at high $p_T$~\cite{Dong:2019byy}. The present work focuses on low-$p_T$ diffusion, specifically the spatial diffusion coefficient $D_s$ of open charm mesons.

Ideally, the transport of heavy flavor in the medium is described by the relativistic Boltzmann transport equation. In the limit where the heavy-flavor mass (and typical momentum) is much larger than the temperature of the surrounding medium --- such that the momentum transfer per collision is small compared to the heavy-quark momentum --- the Boltzmann equation reduces to a Fokker-Planck equation and its stochastically equivalent Langevin equation~\cite{Rapp:2018qla}. This is particularly valid for heavy hadrons -- including $D$ and $B$ mesons and $\Lambda_c$ baryons -- in the hadronic medium where they are not far from equilibrium~\cite{Tolos:2016slr,Das:2024vac} which is the focus of our present study. Traditionally, perturbative calculations, (since heavy flavor mass is much larger that the typical QCD scale $\Lambda_{QCD}$ \cite{Zhao:2020jqu}) including pQCD techniques with Hard Thermal Loop (HTL) corrections \cite{Svetitsky:1987gq,Braaten:1991jj,Braaten:1991we,Moore:2004tg}, were employed to obtain the inputs for the collision kernel, including the transition rates, in the weak-coupling regime at high temperatures. In contrast, non-perturbative calculations, such as lattice QCD simulations \cite{Kaczmarek:1999mm,Banerjee:2011ra,Bazavov:2014kva,Altenkort:2023oms}, AdS/CFT conjectures~\cite{Casalderrey-Solana:2006fio,Gubser:2006nz} and T-matrix methods \cite{Mannarelli:2005pz,vanHees:2007me,Riek:2010fk,Liu:2017qah}, are used in the strong-coupling regime near the pseudo-critical temperature  $T_c$. 

Several model calculations investigating these heavy quark transport coefficients~\cite{Svetitsky:1987gq,Moore:2004tg,Gossiaux:2009hr,vanHees:2007me,Banerjee:2011ra,Riek:2010fk,PhysRevC.82.014908,Banerjee:2011ra,He:2011yi,Ghosh:2011bw,Tolos:2013kva,Ozvenchuk:2014rpa,Berrehrah:2014kba,Berrehrah:2014tva,Tolos:2016slr,Scardina:2017ipo,Torres-Rincon:2021yga,Goswami:2022szb,Goswami:2023hdl,Pooja:2023gqt,Altenkort:2023oms,Satapathy:2022xdw,Dwibedi:2024amt,Dwibedi:2025fnz} and their phenomenological signatures including nuclear suppression factor ($R_{AA}$)~\cite{Gossiaux:2009hr,PhysRevC.82.014908,Moore:2004tg,vanHees:2007me,Ozvenchuk:2014rpa,Scardina:2017ipo} and elliptic flow ($v_2$)~\cite{Gossiaux:2009hr,Moore:2004tg,vanHees:2007me,Ozvenchuk:2014rpa,Scardina:2017ipo} can be found in the literature. The anisotropic diffusion of charm quarks and $D$ mesons in the presence of a background magnetic field \cite{Fukushima:2015wck,Satapathy:2022xdw}, as well as in the case of a rotating fireball \cite{Dwibedi:2024amt,Dwibedi:2025fnz} is also investigated. A substantial portion of the theoretical and computational work in this field has been driven by heavy-ion collision experiments conducted at high energies at the Large Hadron Collider (LHC) at CERN and the Relativistic Heavy Ion Collider (RHIC) at Brookhaven National Laboratory (BNL), both producing nearly baryonless quark-gluon plasma. With the upcoming Compressed Baryonic Matter experiment at FAIR~\cite{Friman:2011zz} and the proposed NA60+ at CERN SPS~\cite{NA60:2022sze,Arnaldi:2025ikz}, heavy-ion collisions will be explored at lower temperatures and high baryon densities. These conditions are expected to enhance the sensitivity to transport and hadronisation properties of charm quarks near the pseudo-critical temperature and in finite-baryon-density matter, leading to renewed interest in the study of QCD matter at finite densities~\cite{Bazavov:2017dus,Elfner:2022iae}.  A comprehensive review of heavy flavor diffusion at small net baryon densities can be found in~\cite{Das:2024vac} while ~\cite{Hosaka:2016ypm,Montana:2023sft} reviews diffusion in nuclear matter. The drag and diffusion coefficients of $D$ mesons at finite baryon density is studied in~\cite{Tolos:2013kva,Ozvenchuk:2014rpa} within a unitarized approach based on chiral and heavy quark spin symmetric effective models.

In this paper, we employ the framework of kinetic theory in the relaxation time approximation in line with earlier refs. \cite{Satapathy:2022xdw,Dwibedi:2024amt,Dwibedi:2025fnz} to obtain the spatial diffusion coefficient of $D$ meson similar to other transport coefficients owing to the fact that the charm quark and the $D$ meson thermalizes within the fireball lifetime \cite{Francis:2015daa,Schlichting:2019abc,Capellino:2022nvf}. We make use of the Einstein's diffusion relation that connects the spatial diffusion coefficient of the particle with its conductivity and susceptibility. The chiral SU(3) hadronic model is adopted to model the baryon-rich background and probe the in-medium modifications of $D$ mesons in dense nuclear matter. The SU(3) chiral effective model is a non-perturbative framework capable of describing strongly interacting matter at finite baryon densities, including finite nuclei, neutron stars, and hyperonic matter \cite{Holt:2014hma,Papazoglou:1997uw, Papazoglou:1998vr,Mishra:2003tr, Zschiesche:2003qq, Mishra:2006wy}. The model has been recently used to obtain the thermodynamic and transport properties of matter at finite densities \cite{Marattukalam:2024mef,Marattukalam:2025lpi,Kumar:2025rxj,Rai:2025lxw}. The $D$ meson masses in the hadronic medium are studied in the past in great detail using the chiral effective theory \cite{Mishra:2003se, Kumar:2010gb}, PNJL model \cite{Blaschke:2011yv}, QCD sum rules \cite{Hayashigaki:2000es, Wang:2015uya, Kumar:2019axp} and coupled channel approach~\cite{Tolos:2004yg,Tolos:2005ft, Mizutani:2006vq,Tolos:2009nn}. In the hadronic domain as we move to higher densities the transport properties are no longer pion dominated rather the heavier baryons including the nucleons begin to dominate the medium properties \cite{Tolos:2013kva}. Since we are concerned of the $D$ meson diffusion in hadronic matter here, the major contribution to transport coefficients can be attributed to scattering of $D$ meson off the nucleons. The hadronic chiral effective model is well equipped to explain matter in this domain. 

The paper is organized as follows. In Section~\ref{sec:form1}, we outline the theoretical framework describing the diffusion of $D$ mesons and the evaluation of the diffusion coefficients. The chiral SU(3) hadronic model is discussed in Section~\ref{sec:form2}. The diffusion of $D$ mesons through dense nuclear matter and the corresponding results are presented in Section~\ref{sec:results}, followed by a summary in Section~\ref{summary}.

\section{Formalism}
\label{sec:formalism}

\subsection{$D$ meson diffusion coefficients}\label{sec:form1}
The time scale of heavy-flavor transport is much larger than that of light quarks, and it is well established that the macroscopic evolution of heavy flavor can be modeled using Langevin simulations \cite{Svetitsky:1987gq,Moore:2004tg,Petreczky:2005nh}, in which the position and momentum of the heavy flavor evolve (non-relativistically since $M\gg T$) according to stochastic differential equations, 
\begin{align}
	\frac{dx^i}{dt}&=\frac{p^i}{M},\\
	\frac{dp^i}{dt}&=-\eta_D^{ij} p^j+\xi^i(t),
\end{align}
where $x^i$, $p^i$, $M$ are the position, momentum and mass of the heavy flavor, $\eta_D^{ij}$ specifies the (momentum) drag and the stochastic forces satisfy,
\begin{align}
	\langle\xi^i(t)\rangle&=0\\
	\langle\xi^i(t)\xi^j(t')\rangle&=2B^{ij}\delta(t-t'),
\end{align}
with $B^{ij}$ being the (momentum) diffusion coefficients. While the momentum drag and diffusion coefficients serve as inputs for transport simulations, one can also describe the system through the spatial diffusion coefficients, $D_s^{ij}$ defined in terms of the mean square displacement \cite{Moore:2004tg,Petreczky:2005nh,Capellino:2022nvf}. For the (heavy) Brownian particle, the mean square displacement is $\langle x^2 \rangle = \sigma^2(t) \approx 6 D_s t$, with $\langle x^i(t) x^j(t) \rangle = 2 D_s t\, \delta^{ij}$, for a particle following isotropic diffusion governed by the Fick's law,
\begin{equation}
	\partial_t n = D_s \nabla^2 n,
\end{equation}
where $D_s^{ij} = D_s \delta^{ij}$. Here, $D_s$ is the isotropic spatial diffusion coefficient and $n = n(\vec{x}, t)$ is the number density profile in space and time. 
For an isotropic medium where $\eta_D^{ij}=\gamma\delta^{ij}$, $B^{ij}=D_p\delta^{ij}$ the momentum drag $(\gamma)$, momentum diffusion $(D_p)$ and spatial diffusion $(D_s)$ coefficients are connected to each other through the following relations $D_s=\frac{T}{M\gamma}$, $D_s=\frac{T^2}{D_p}$ and $\gamma=\frac{D_p}{MT}$.

In the relativistic generalization the fluctuation dissipation relation takes the form \cite{Moore:2004tg}
\begin{align}
	\gamma=\frac{D_p}{ET},
\end{align}
where $E=\sqrt{p^2+M^2}$ is the energy of the particle and the use of Langevin approach is useful only in scenarios where the momentum transfer is small satisfying this relation. Also, the relation $D_s=\frac{T^2}{D_p}$ arises naturally and holds even in the relativistic case as shown in \cite{Capellino:2022nvf}, also giving $D_s=\frac{T}{E\gamma}$.

Under It\^{o} discretization the relativistic Langevin approach in the above form can be equivalently described through the Fokker-Planck approach as
\begin{equation}
	p^\mu\frac{\partial f}{\partial x^\mu}=p^0\frac{\partial}{\partial p^i}\left[A^i f+\frac{\partial}{\partial p^j}B^{ij} f\right],
\end{equation}
where $A^i=\eta_D^{ij}p^j$, $p^0=E$ and $f$ is the distribution function of the incoming heavy flavor.

Here, we consider the diffusion of charmed mesons -- $D$ mesons -- through a hadronic (nuclear) medium. The spatial diffusion $D_s$ of the $D$ meson is related to its conductivity $\sigma_D$ and susceptibility $\chi_D$ through the relation ~\cite{Romatschke:2017ejr,Riek:2010py}
\begin{equation}
	D_s = \frac{\sigma_D}{\chi_D}.\label{D_s}
\end{equation}In the framework of kinetic theory, the macroscopic and microscopic expressions of dissipative current density are given as ~\cite{Satapathy:2022xdw,Dwibedi:2024amt,Dwibedi:2025fnz} 
\begin{equation}
	J_D^i=\sigma_D^{ij}\nabla_j \mu_D=g\int\frac{d^3p}{(\pi)^3}\frac{p^i}{E_D}\delta f,
\end{equation}
where $\sigma_D^{ij}$ is the conductivity tensor, $\mu_D$ is the chemical potential of the $D$ meson, the gradient of which gives the current density, $g$ is the degeneracy of the $D$ meson and $\delta f$ is the deviation from the equilibrium distribution function $f_0=1/[e^{(E_D-\mu_D)/T}-1]$ where $E_D=\sqrt{p^2+m_D^2}$ with $m_D$ being the mass of the $D$ meson. We employ the relativistic Boltzmann equation (RBE) in the relaxation time approximation to obtain the form of $\delta f$, and comparing the microscopic expression with the macroscopic expression, we can obtain the expression for $D$ meson conductivity in an isotropic medium as~\cite{Satapathy:2022xdw,Dwibedi:2024amt,Dwibedi:2025fnz} 	
\begin{equation}
	\sigma_D = \frac{1}{3T}\int\frac{d^3p}{(2\pi)^3}\frac{p^2}{E_D^2}\tau_c~ f_0\big(1 + f_0\big),\label{sigma}
\end{equation}
where $\tau_c$ is the relaxation time of the $D$ meson in the medium. The expression of $D$ meson susceptibility is given by~\cite{Satapathy:2022xdw,Dwibedi:2024amt,Dwibedi:2025fnz} 
\begin{equation}
	\chi_D = \frac{\partial n_D}{\partial \mu_D} = \frac{1}{T}\int\frac{d^3p}{(2\pi)^3}f_0(1+f_0),\label{chi}
\end{equation}
where $n_D$ is the number density of the $D$ mesons in the medium. Reader can look into Refs.~\cite{Satapathy:2022xdw,Dwibedi:2024amt,Dwibedi:2025fnz} for a detailed discussion.

\subsection{Chiral SU(3) hadronic model}\label{sec:form2}
We adopt the chiral SU(3) mean field hadronic model to study hadron-hadron interaction at finite temperature and density, which depends on a nonlinear realization of chiral symmetry and incorporates the effects of broken scale invariance~\cite{Weinberg:1968de,Bardeen:1969ra,Papazoglou:1998vr,Hayano:2008vn, Kumar:2019axp}. The interaction among hadrons is mediated by the exchange of mesonic fields.  Specifically, the scalar sector features the $\sigma$ field, which is associated with the non-strange scalar meson having quark content $\bar{u}u+\bar{d}d$, the $\zeta$ field, which is composed of strange quark content $\bar{s}s$ and to address isospin-asymmetric matter, the scalar-isovector $\delta$ field, with quark content $\bar{u}u-\bar{d}d$, is utilized. The vector sector is defined by the non-strange $\omega$ and the isospin-sensitive $\rho$ fields. The fundamental effective Lagrangian density for the chiral SU(3) model is written as
\begin{equation}
\label{gen_L}
\mathcal{L} 
  = \mathcal{L}_{\text{kin}} + \mathcal{L}_{\text{BM}} +
	\mathcal{L}_{\text{scal}}+ \mathcal{L}_{\text{vec}} + \mathcal{L}_{\text{SB}}.  
\end{equation}
In Eq. (\ref{gen_L}), the kinetic energy term for various particles is represented by $\mathcal{L}_{\text{kin}}$, whereas the interaction of baryons with different mesons is denoted by $\mathcal{L}_{\text{BM}}$. The self-interaction terms of scalar and vector mesons is described by the third term $\mathcal{L}_{\text{scal}}$ and fourth term $\mathcal{L}_{\text{vec}}$ of above Lagrangian density, respectively. The final term  $\mathcal{L}_{\text{SB}}$ represents to the explicit breaking of chiral symmetry. For more detailed explanations of the interaction Lagrangian density terms, refer to the literature \cite{Papazoglou:1998vr,Cruz-Camacho:2024odu, Kaur:2024cfm}. 
In this work, we employ the mean-field approximation, treating all mesonic fields as the classical fields. Consequently, only the vector and scalar fields contribute to the Lagrangian term of the nucleon-meson interaction, as the expectation values of the other mesons are zero. For a strongly interacting baryonic system, the partition function within the grand canonical ensemble is given by \cite{Papazoglou:1998vr,Zschiesche:1999gf, Kaur:2024cfm} 
\begin{equation}
\mathcal{Z} = \text{Tr} \exp [- (\hat{\mathcal{H}} - \sum_{N} \mu_N \hat{\mathcal{H}}_N)/T],
\end{equation}
where $\mu_N$ stands for the chemical potential of nucleons ($N=p,n$) and $\hat{\mathcal{H}}$, $\hat{\mathcal{N}}_N$ are the operators of the Hamiltonian and number density, respectively. The definition of the thermodynamic potential, $\Omega$, at a given temperature $T$ is \cite{Zschiesche:1999gf}
\begin{equation}
\Omega(T, V, \mu) = -T \ln \mathcal{Z}.
\end{equation}
The thermodynamic potential per unit volume $\frac{\Omega}{V}$, can be written as follows by substituting mean-field Hamiltonian density in the form of Lagrangian density \cite{Zschiesche:2003qq}
\begin{align}
\frac{\Omega}{V}
&= -\frac{g_{N} T}{(2\pi)^3}\sum_{N=p,n}  \int d^3p
\Big[\ln\left(1+e^{-\left[E_N^*(p)-\mu_N^*\right]/T}\right)\nonumber\\
&+\ln\left(1+e^{-\left[E_N^*(p)+\mu_N^*\right]/T}\right)
\Big]- \mathcal{L}_{v e c}-\mathcal{L}_0-\mathcal{L}_{S B} - \mathcal{V}_{\text{vac}}.
\label{Eq_TP}
\end{align}
To obtain a zero vacuum energy at $\rho_B=0$, the vacuum energy $\mathcal{V}_\text{vac}$ has been subtracted in Eq.~(\ref{Eq_TP}). The factor $g_N$ signify the spin-isospin degeneracy factor, which equals $2$ for nuclear matter. The energy of a single particle within a medium is expressed as \(E_N^*(p) = \sqrt{p^2 + M_N^{*2}}\). The effective baryon mass is determined by the equation 
\begin{equation}
	M_N^* = -\left(g_{\sigma N} \sigma + g_{\zeta N} \zeta + g_{\delta N} \tau_{3N} \delta\right),
\end{equation}
where \(g_{\sigma N}\), \(g_{\zeta N}\), and \(g_{\delta N}\) represent the coupling constants of the \(\sigma\), \(\zeta\), and \(\delta\) fields, respectively, with nucleons. Additionally, \(\tau_{3N}\) takes the value of \(+1\) for \(p\) and \(-1\) for \(n\). The effective chemical potential is defined as 
\begin{equation}
	\mu_N^* = \mu_N - g_{\omega N} \omega- g_{\rho N} \tau_{3N} \rho,
\end{equation}
where \(g_{\omega N}\) and \(g_{\rho N}\) serve as the coupling constants of the nucleons with \(\omega\), and \(\rho\) vector fields, respectively. 
The values of the mesonic fields are obtained by extremizing the thermodynamic potential, the coupled equations of motion for mesonic field can be expressed in the following relations   
\begin{align}
\label{Eq_sigma_eq1}
\frac{\partial (\Omega/V)}{\partial \sigma}& =  k_0\chi^2\sigma-4 k_1\left(\sigma^2+\zeta^2+\delta^2\right) \sigma-2 k_2\left(\sigma^3+3 \sigma \delta^2\right) 
 \nonumber\\ &-2 k_3 \chi \sigma \zeta -\frac{d}{3} \chi^4\left(\frac{2 \sigma}{\sigma^2-\delta^2}\right)+\frac{\chi^2}{\chi_0^2} m_\pi^2 f_\pi \nonumber\\ & 
 -\sum_{N= p, n} g_{\sigma N} \rho_N^s=0 ,
\end{align}
\begin{align}
\frac{\partial  (\Omega/V)}{\partial \zeta}& =  k_{0}\chi^{2}\zeta-4k_{1}\left( \sigma^{2}+\zeta^{2}+\delta^{2}\right)
\zeta-4k_{2}\zeta^{3}\nonumber\\
&-k_{3}\chi\left( \sigma^{2}-\delta^{2}\right) 
-\frac{d}{3}\frac{\chi^{4}}{\zeta} +\left(\frac{\chi}{\chi_{0}} \right)
^{2}\nonumber\\
&\left[ \sqrt{2}m_{K}^{2}f_{K}-\frac{1}{\sqrt{2}} m_{\pi}^{2}f_{\pi}\right]-\sum_{N= p, n} g_{\zeta N}\rho_{N}^{s}=0 ,
 \label{Eq_zeta11}
\end{align}
\begin{align}
\label{Eq_delta}
	\frac{\partial  (\Omega/V)}{\partial \delta}& = k_0 \chi^2 \delta-4 k_1\left(\sigma^2+\zeta^2+\delta^2\right) \delta-2 k_2\left(\delta^3+3 \delta \sigma^2\right)
 \nonumber\\
&-2 k_3 \chi \delta \zeta-\frac{d}{3} \chi^4\left(\frac{2 \delta}{\sigma^2-\delta^2}\right)   
	 \nonumber\\
	&-\sum_{N= p, n} g_{\delta N}I_{3N} \rho_N^s=0 ,
\end{align}
\begin{align}
\label{Eq_chi}
\frac{\partial  (\Omega/V)}{\partial \chi}& =  k_0 \chi\left(\sigma^2+\right.\left.\zeta^2+\delta^2\right)-k_3 \chi\left(\sigma^2-\delta^2\right) \zeta\nonumber\\ &+\left[4 k_4-d+1+\ln \frac{\chi^4}{\chi_0^4} -\frac{4 d}{3} \ln \left(\frac{\left(\sigma^2-\delta^2\right) \zeta}{\sigma_0^2 \zeta_0}\right)\right]\nonumber\\ &\chi^3 + \frac{2 \chi}{\chi_0^2}\left[m_\pi^2 f_\pi \sigma+\left(\sqrt{2} m_K^2 f_K-\frac{1}{\sqrt{2}} m_\pi^2 f_\pi\right) \zeta\right]\nonumber\\ & -\frac{\chi}{\chi_0^2}\left(m_\omega^2 \omega^2+m_\rho^2 \rho^2\right)=0 ,
\end{align}

\begin{align}
\label{Eq_omega_field}
	\frac{\partial  (\Omega/V)}{\partial \omega}& = \frac{\chi^2}{\chi_0^2}\left(m_\omega^2 \omega\right)+4 g_4 \omega^3+12 g_4 \omega \rho^2 \nonumber\\
	&-\sum_{N= p, n} g_{\omega N} \rho_N^v=0 ,
\end{align}
\begin{align}
\label{Eq_rho_field}
	\frac{\partial (\Omega/V)}{\partial \rho}& = \frac{\chi^2}{\chi_0^2}\left(m_\rho^2 \rho\right)+4 g_4 \rho^3+12 g_4 \omega^2 \rho \nonumber\\
	&-\sum_{N= p, n} g_{\rho N}I_{3N} \rho_N^v= 0 .
\end{align}
By solving the nonlinear system of Eqs.~(\ref{Eq_sigma_eq1}) to (\ref{Eq_rho_field}), we analyze the properties of baryonic medium at various values of temperature $T$, baryonic density $\rho_B$, and isospin asymmetry $I_a$ of medium, which is defined as $-\frac{\sum_{N} I_{3N} \rho^{v}_{N}}{\rho_{B}}$ with $I_{3N}$ being the third component of isospin of nucleons. Moreover, $\rho_N^v$ and $\rho_N^s$ are vector and scalar densities of the baryons, respectively, defined as follows
 \begin{align}
\rho_{N}^{v} = g_{N}\int\frac{d^{3}p}{(2\pi)^{3}}  
\Bigg(\frac{1}{1+\exp\left[(E^{\ast}_N(p) 
-\mu^{*}_{N})/T \right]}\nonumber\\
-\frac{1}{1+\exp\left[(E^{\ast}_N(p)
+\mu^{*}_{N})/T \right]}\Bigg) ,
\label{rhov0}
\end{align}
and
\begin{align}
\rho_{N}^{s} = g_{N}\int\frac{d^{3}p}{(2\pi)^{3}} 
\frac{m_{N}^{*}}{E^{\ast}_i(p)} \Bigg(\frac{1}{1+\exp\left[(E^{\ast}_N(p) 
-\mu^{*}_{N})/T \right]}\nonumber\\
+\frac{1}{1+\exp\left[(E^{\ast}_N(p)
+\mu^{*}_{N})/T \right]}\Bigg) .
\label{rhos0}
\end{align}
Table \ref{table_chiral_para} lists the important parameters employed in this work, such as meson vacuum masses and mesonic field coupling constants with nucleons.

\begin{table}[h]
\centering
\caption{ The coupling constants of mesonic fields with nucleons and different parameters utilized in the current calculations.}
\begin{tabular}{|l|l|l|}
\hline
$k_0 = 2.53$ & $\sigma_0 = -93.3$ MeV & $g_{\sigma N} = 10.6$ \\ \hline
$k_1 = 1.35$ & $\zeta_0 = -106.76$ MeV & $g_{\zeta N} = -0.46$ \\ \hline
$k_2 = -4.77$ & $\chi_0 = 409.76$ MeV & $g_{\delta N} = 2.34$ \\ \hline
$k_3 = -2.77$ & $d = 0.064$ & $g_{\omega N} = 13.42$ \\ \hline
$k_4 = -0.21$ & $g_4 = 79.9$ & $g_{\rho N} = 5.48$ \\ \hline
$m_{\pi} = 139$ MeV & $m_K = 498$ MeV & $f_{\pi} = 93.3$ MeV \\ \hline
$f_K = 122.14$ MeV & $\rho_0 = 0.15$ fm$^{-3}$ & $m_N = 939$ MeV \\ \hline
$m_{\omega} = 783$ MeV & $m_{\rho} = 770$ MeV & $f_D = 135$ MeV \\ \hline
\end{tabular}
\label{table_chiral_para}
\end{table}

Next, we derive the dispersion relation for $D$ and $\bar D $ mesons within isospin asymmetric nuclear medium. Under the chiral SU(3) framework, the medium-modified energies of $D$ and $\bar D $ mesons arise from their interactions with nucleons and the scalar meson fields. The Lagrangian density describing these interactions can be expressed as \cite{Mishra:2008dj,Kumar:2010gb, Kaur:2024cfm, Kaur:2025kjk} 
\begin{widetext}
\begin{align}
	\mathcal{L}_{DN}
	=& -\frac{i}{8f_D^2}\Bigg[
	3\left(\bar p\gamma^\mu p+\bar n\gamma^\mu n\right)
	\Big\{
	D^0(\partial_\mu \bar D^0)-(\partial_\mu D^0)\bar D^0+ D^+(\partial_\mu D^-)-(\partial_\mu D^+)D^-
	\Big\}+\left(\bar p\gamma^\mu p-\bar n\gamma^\mu n\right)
	\Big\{
	D^0(\partial_\mu \bar D^0)	\nonumber\\
	&
	-(\partial_\mu D^0)\bar D^0-\Big(D^+(\partial_\mu D^-)-(\partial_\mu D^+)D^-\Big)
	\Big\}
	\Bigg]+\frac{m_D^2}{2f_D}\Bigg[
	(\sigma+\sqrt2\,\zeta_c)\Big(\bar D^0D^0 + D^-D^+\Big)+\delta\Big(\bar D^0D^0 - D^-D^+\Big)
	\Bigg]
	\nonumber\\
	&-\frac{1}{f_D}\Bigg[
	(\sigma+\sqrt2\,\zeta_c)
	\Big(
	(\partial_\mu \bar D^0)(\partial^\mu D^0)+(\partial_\mu D^-)(\partial^\mu D^+)
	\Big)+\delta
	\Big(
	(\partial_\mu \bar D^0)(\partial^\mu D^0)
	-(\partial_\mu D^-)(\partial^\mu D^+)
	\Big)
	\Bigg]	\nonumber\\
	&+\frac{d_1}{2f_D^2}\left(\bar pp+\bar nn\right)
	\Big(
	(\partial_\mu D^-)(\partial^\mu D^+)+(\partial_\mu \bar D^0)(\partial^\mu D^0)
	\Big)+\frac{d_2}{4f_D^2}\Bigg[
	(\bar pp+\bar nn)
	\Big(
	(\partial_\mu \bar D^0)(\partial^\mu D^0)	+(\partial_\mu D^-)(\partial^\mu D^+)
	\Big)	\nonumber\\
	&
	+(\bar pp-\bar nn)
	\Big(
	(\partial_\mu \bar D^0)(\partial^\mu D^0)-(\partial_\mu D^-)(\partial^\mu D^+)
	\Big)
	\Bigg].
	\label{Lang_DN}
\end{align}
\end{widetext}
In the above equation, the first term corresponds to the Weinberg-Tomozawa vectorial interaction which is extracted from kinetic component of Eq. (\ref{gen_L}) while the second term accounts for explicit symmetry breaking which gives attractive interaction for $D$ mesons, referred to as the scalar meson exchange term and the third term is generated by kinetic term of the pseudoscalar meson of interaction Lagrangian, identified as 1st range term\cite{Mishra:2008kg, Mishra:2006wy}. Lastly, the fourth and fifth terms represent isospin-dependent range terms derived from the baryon-meson interaction Lagrangian of this model \cite{Mishra:2006wy, Mishra:2004te}. The parameters \(d_1\) and \(d_2\), in the above equation, are assigned values of 2.56/$m_K$ and 0.73/$m_K$, respectively, based on empirical data for the kaon-nucleon scattering length \cite{Barnes:1992ca}. The medium-modified masses of $D$ ($D^0,D^+$)  and $\bar D$ ($\bar D^0,D^-$) mesons are calculated from Eq. (\ref{Lang_DN}) via Euler-Lagrange equation of motion and Fourier transformation to obtain the dispersion relation,
\begin{equation}
	E_D^{*2}+ {\vec p_D}^2 + m_D^2 -\Pi^*(E_D^*, |\vec p_D|)=0.
	\label {dispdm}
\end{equation}
Here, \(E_D^* = \sqrt{p_D^2 + m_D^{*2}}\) and $\Pi^*(E_D^*, |\vec p_D|)$ denote the in-medium energy and self energy for $D$ meson, respectively, with analogous definitions applying to the $\bar D$ meson. The parameter \( m_D \) represents the vacuum mass of the \( D(\bar{D}) \) meson, with values of 1869 MeV for \( D^{\pm} \) mesons and 1864.5 MeV for \( D^0 \) and \( \bar{D}^0 \) mesons. Explicitly, $\Pi^*(E_D^*, |\vec p_D|)$ for the $D$ meson doublet is given as
\begin{eqnarray}
	\Pi^*(E_D^*, |\vec p_D|) &= & \frac {1}{4 f_D^2}\Big [3 (\rho_p^v +\rho_n^v)
	\pm (\rho_p^v -\rho_n^v) \big)
	\Big ] E_D^* \nonumber \\
	&+&\frac {m_D^2}{2 f_D} (\sigma ' +\sqrt 2 {\zeta_c} ' \pm \delta ')
	\nonumber \\ & +& \Big [- \frac {1}{f_D}
	(\sigma ' +\sqrt 2 {\zeta_c} ' \pm \delta ')
	+\frac {d_1}{2 f_D ^2} (\rho_p ^s +\rho_p ^s)\nonumber \\
	&+&\frac {d_2}{4 f_D ^2} \Big ((\rho^s_p +\rho^s_n)
	\pm   (\rho^s_p -\rho^s_n) \Big ) \Big ]\nonumber \\
	&&(E_D^{*2} - {\vec p_D}^2).
	\label{self_D}
\end{eqnarray}
The $\pm$ sign correspond to $D^0$ and $D^+$ mesons respectively. The deviation of the $\sigma$, $\zeta_c$, and $\delta$ fields from their respective vacuum expectation values $\sigma_0$, $\zeta_{c 0}$, and $\delta_0$, are represented by the symbols $\sigma '$, $\zeta_c '$, and $\delta '$. For the present calculation we set $\delta_0 = 0$ to preserve the isospin symmetry of the vacuum and the interaction of scalar quark condensate  ($\zeta_c$) induces only minor mass modifications and hence its medium deviations are neglected \cite{Roder:2003uz,Kumar:2010gb}. The proton and neutron number densities in the above Eq. (\ref{self_D}) are $\rho_p^v$ and $\rho_n^v$, and their scalar densities are $\rho^s_p$ and $\rho^s_n$, which are obtained by solving Eqs.~(\ref{Eq_sigma_eq1}) to (\ref{Eq_rho_field}).   
Likewise, the self-energy corresponding to $\bar D$ meson doublet, (${\bar D}^0$,$D^-$) is expressed as
\begin{eqnarray}
	\Pi^*(E_{\bar D}^*, |\vec p_D|) &= &- \frac {1}{4 f_D^2}\Big [3 (\rho_p^v +\rho_n^v)
	\pm (\rho_p^v -\rho_n^v) \big)
	\Big ] E_{\bar D}^* \nonumber \\
	&+&\frac {m_D^2}{2 f_D} (\sigma ' +\sqrt 2 {\zeta_c} ' \pm \delta ')
	\nonumber \\ & +& \Big [- \frac {1}{f_D}
	(\sigma ' +\sqrt 2 {\zeta_c} ' \pm \delta ')
	+\frac {d_1}{2 f_D ^2} (\rho_p ^s +\rho_p ^s)\nonumber \\
	&+&\frac {d_2}{4 f_D ^2} \Big ((\rho^s_p +\rho^s_n)
	\pm   (\rho^s_p -\rho^s_n) \Big ) \Big ]\nonumber \\
	&&(E_{\bar D}^{*2} - {\vec p_D}^2),
	\label{self_bD}
\end{eqnarray}
here, the $\pm$ signs are associated with the $\bar {D^0}$ and $D^-$ respectively.

\section{Results and discussion}
\label{sec:results}

We investigate the transport properties of open-charm $D$ mesons
($D^0$, $D^+$, $\bar{D}^0$, $D^-$) propagating through dense
nuclear matter at finite temperature and isospin asymmetry as function of scaled baryon density $\rho_B/\rho_0$, where $\rho_0$ denotes the nuclear matter saturation density ($\rho_0$ = 0.15 $\text{fm}^{-3}$).
Using the chiral SU(3) hadronic model along with the kinetic theory, we compute the in-medium
masses, relaxation times, spatial diffusion coefficients, momentum diffusion coefficients and momentum drag coefficients of $D$ mesons as functions of scaled baryon density
$\rho_B/\rho_0$ for temperatures $T = 20$--$150$ MeV and isospin
asymmetry parameter $I_a = 0$, $0.3$, and $0.5$.

\begin{figure}
	\includegraphics[width=\linewidth]{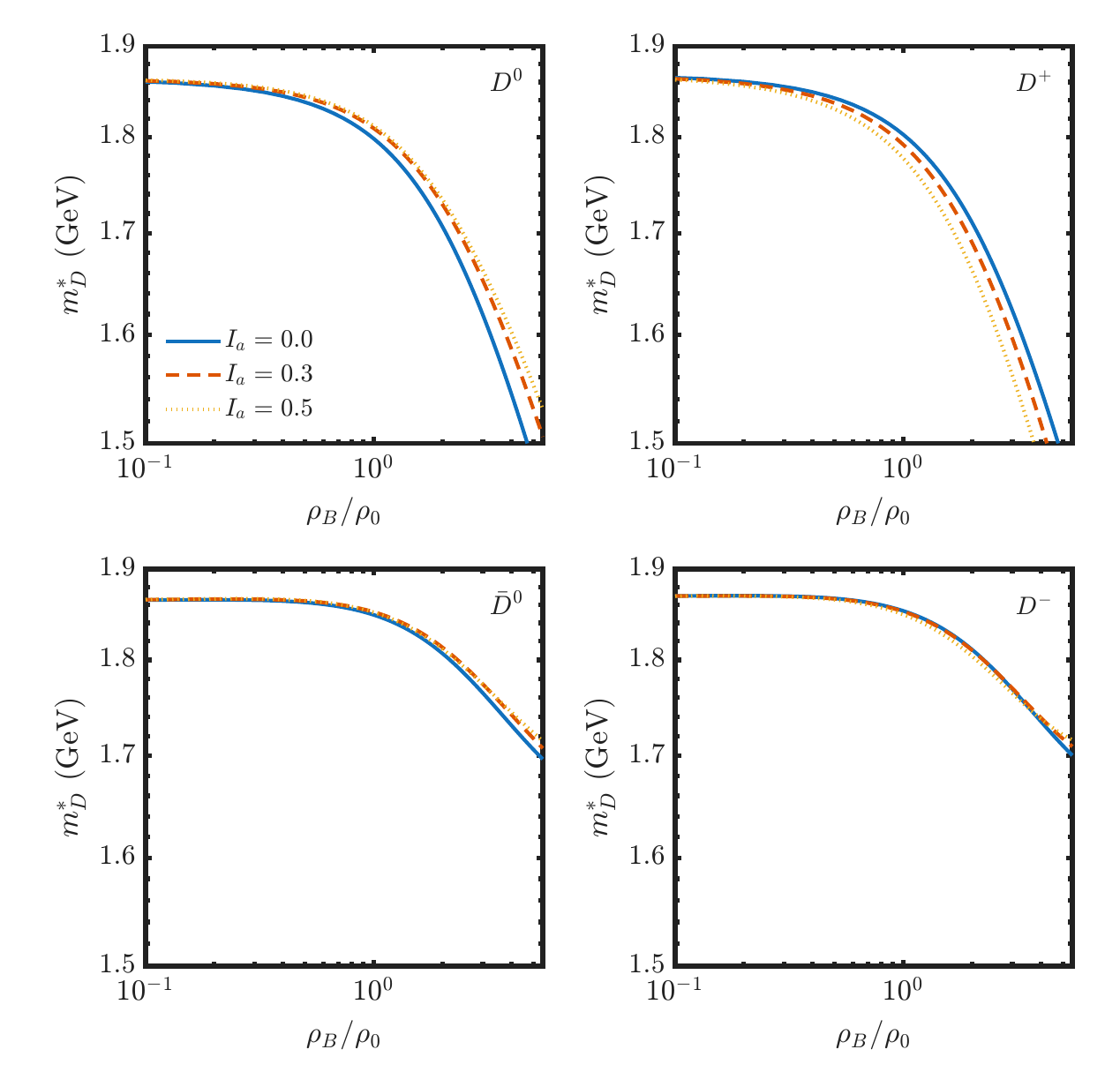}
	\caption{In-medium masses of $D^0$, $D^+$, $\bar{D}^0$, and $D^-$ at momentum $p=0$ as function of scaled baryon density $\rho_B/\rho_0$ at $T = 100$ MeV for three isospin asymmetry values $I_a=0,0.3,0.5.$}
	\label{fig:mass_eta}
\end{figure}

\begin{figure}
	\includegraphics[width=\linewidth]{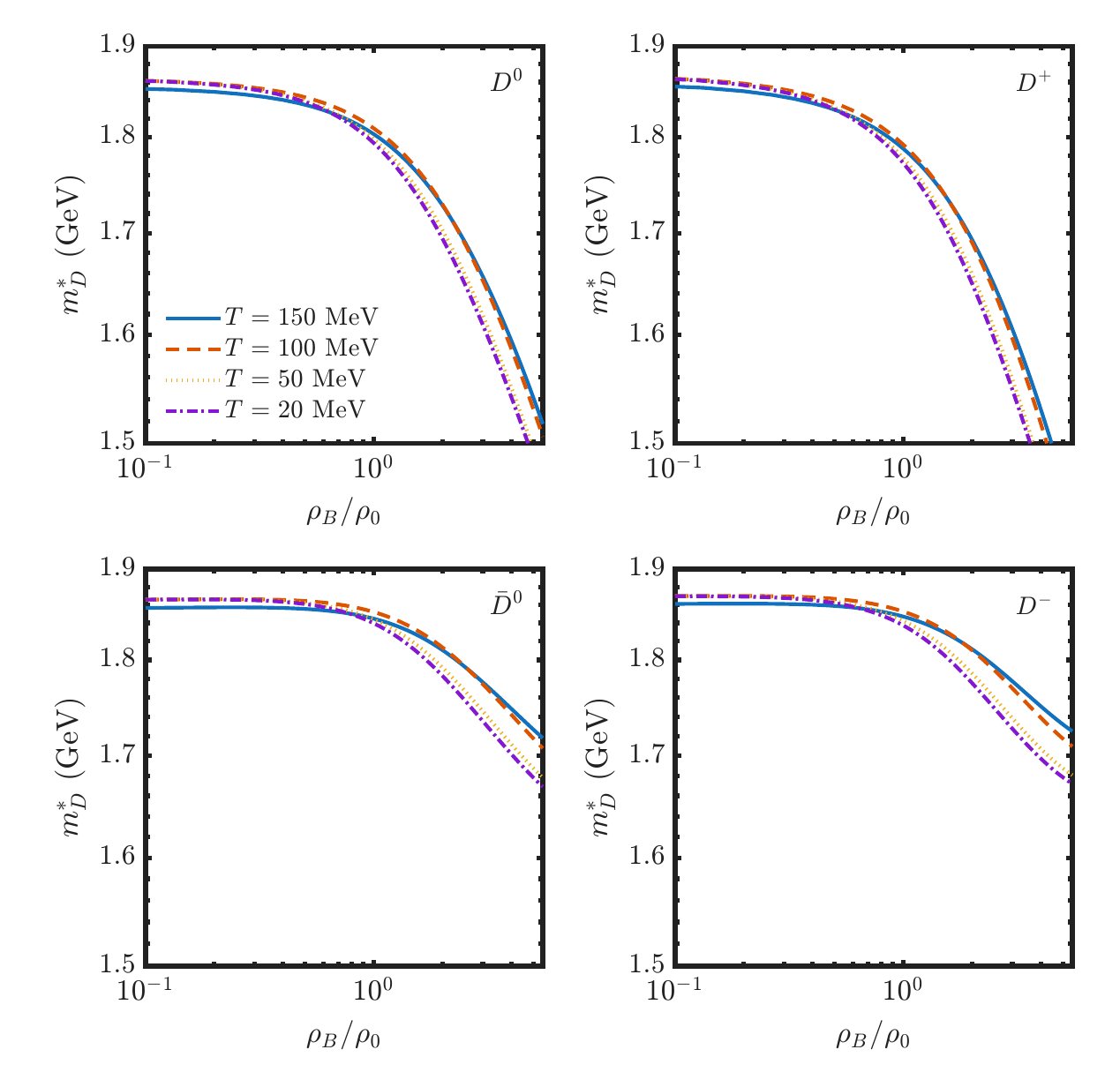}
	\caption{In-medium masses of $D^0$, $D^+$, $\bar{D}^0$, and $D^-$ at momentum $p=0$ as function of scaled baryon density $\rho_B/\rho_0$ for different temperatures  $T = 20,50,100,150$ MeV at isospin asymmetry value $I_a=0.3.$}
	\label{fig:mass_temp}
\end{figure}

Fig.~\ref{fig:mass_eta} presents the in-medium masses of the $D$ mesons, $m^*_D$, as function of scaled baryon density $\rho_B/\rho_0$ at a fixed temperature $T = 100$ MeV for $I_a = 0$ (symmetric nuclear matter),
$0.3$ (asymmetric matter), and $0.5$ (pure neutron matter).  In all four panels, the meson masses remain close to their vacuum values at low densities and exhibit a monotonic decrease with increasing baryon density of the medium, reflecting the attractive nature of the scalar interaction in the dense medium \cite{Kumar:2010gb}. We observe similar decline in the effective masses $m^*_{D^0}$ and  $m^*_{D^+}$ with respect to density for symmetric nuclear matter ($I_a=0$). This reduction is mainly attributed to the dominant attractive influence of the Weinberg-Tomozawa (WT) term, which is a key component in the interaction terms in the Lagrangian of Eq.(\ref{Lang_DN}). Although, scalar meson exchange and range terms provide additional attractive contributions, the behavior of $D$ mesons masses is primarily influenced by WT term \cite{Mishra:2008cd, Kumar:2009xc}.  
However,  as we move from isospin symmetric to asymmetric nuclear matter, a mass splitting of $D^0$ and $D^+$ mesons becomes evident. This asymmetry results in a less pronounced decrease in $m^*_{D^0}$, while the mass of $D^+$ mesons undergoes a more significant reduction, as depicted in Fig.~\ref{fig:mass_eta} (upper panel).
In the lower panel of Fig.~\ref{fig:mass_eta}, the antiparticle channels exhibit a notably smaller reduction in their effective masses ( $m^*_{\bar D}$) with increasing density reaching just around 1.72 GeV from their vacuum value of 1.869 GeV at  $\rho_B/\rho_0 \approx 4$. However, in particle channel, the $m^*_{D}$ values fall to approximately 1.4 GeV from their vacuum value of 1.8645~GeV under same environmental conditions. For the $\bar{D}^0$ meson (lower left panel) and the $D^-$ meson (lower right panel), the three curves are nearly degenerate (in comparison with the particle channels), indicating a reduced effect of isospin composition of the medium on the mass of the $\bar{D}^0$ and $D^-$ mesons. In this scenario, the WT term provides a repulsive contribution for the $\bar{D}^0$ and $D^-$ mesons. However, the overall attractive influence of the scalar and range terms surpasses the repulsive WT term, leading to a significant decrease in $m^*_{\bar D^0}$ and $m^*_{D^-}$  \cite{Mishra:2008cd}.

Fig.~\ref{fig:mass_temp} displays the density dependence of the $D$ ($\bar{D}$) meson effective masses at a fixed isospin asymmetry ($I_a=0.3$) for various temperatures, $T = 20$, 50, 100, and 150 MeV. At low baryon densities, the effect of temperature is visible as a slight downward shift in the effective mass of $D$ and $\bar{D}$ mesons. Whereas, at higher temperatures of medium, the partial restoration of scalar condensates results in a slower reduction of these meson masses, even at moderate densities.
For the $D^0$ meson (upper left panel), the four temperature curves are closely spaced at low densities but begin to fan out beyond $\rho_B/\rho_0 \sim 1$. The lowest temperature ($T = 20$ MeV) yields the steepest descent, where the nuclear medium is denser and the scalar
field takes larger values while the $T = 150$ MeV curve shows the most gradual mass reduction. The $D^+$ channel (upper right panel) again exhibits the largest overall mass drop, and the temperature dependence here is comparatively mild—the four curves remain closely grouped, suggesting that the dominant contribution to the $D^+$ mass shift arises from the density-dependent mean fields rather than from thermal effects. The masses of $\bar{D}^0$ and $D^-$ mesons show a more pronounced temperature splitting at intermediate densities ($1 \lesssim \rho_B/\rho_0 \lesssim 4$) as illustrated in Fig.~\ref{fig:mass_temp}.
\begin{figure}
	\includegraphics[width=0.8\linewidth]{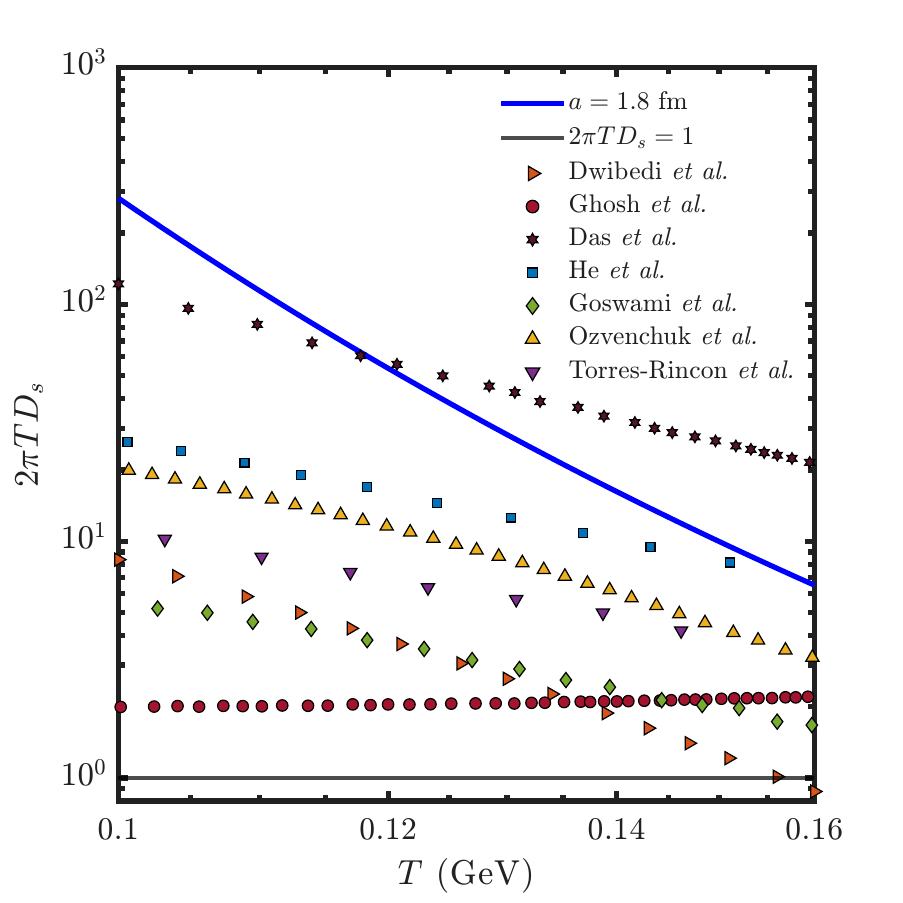}
	\caption{Comparison of scaled spatial diffusion coefficient $2\pi T D_s$ of $D$ mesons through the hadronic medium in heavy-ion collisions as obtained in various studies by He \textit{et al.} \cite{He:2011yi}, Ghosh \textit{et al.} \cite{Ghosh:2011bw}, Das \textit{et al.}~\cite{Das:2011vba}, Ozvenchuk \textit{et al.}~\cite{Ozvenchuk:2014rpa}, Torres-Rincon \textit{et al.} \cite{Torres-Rincon:2021yga},  Goswami \textit{et al.} \cite{Goswami:2023hdl} and Dwibedi \textit{et al.} \cite{Dwibedi:2024amt} plotted as a function of temperature $T$ along with the AdS/CFT bound $2\pi T D_s = 1$. The solid line corresponds to our results for the diffusion through a nuclear medium with average scattering length 1.8 fm.}\label{fig:data}
\end{figure}
\begin{figure*}
	\includegraphics[width=\linewidth]{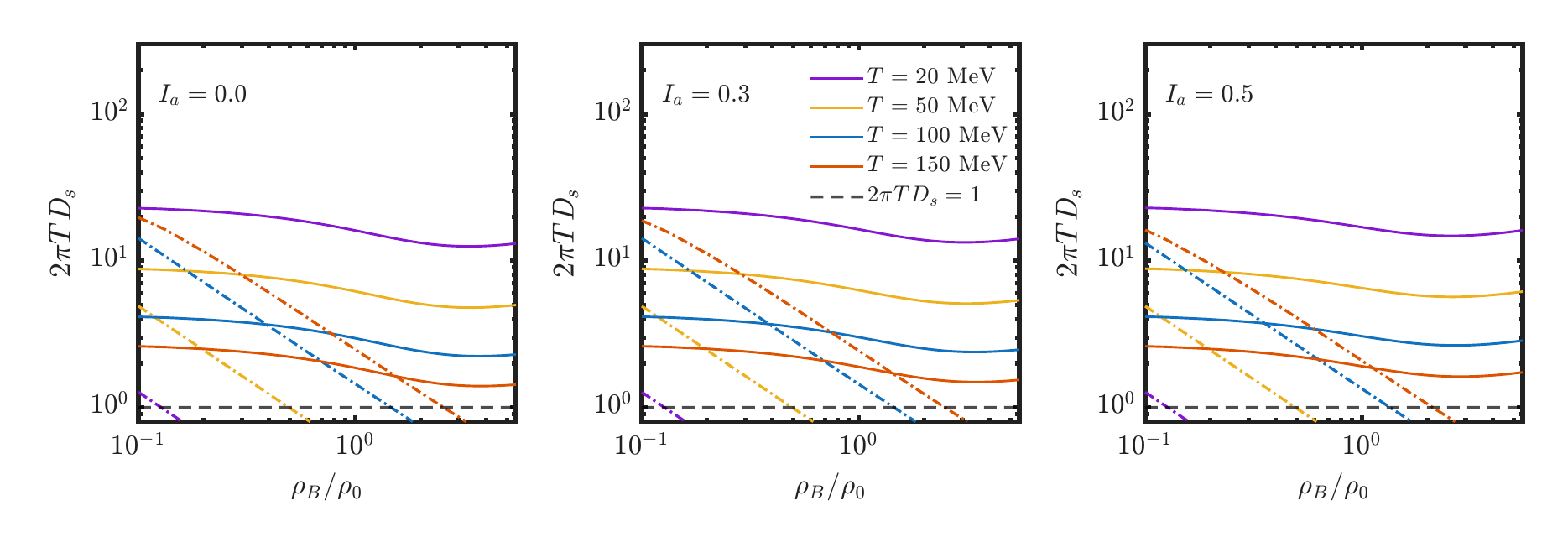}
	\caption{Scaled spatial diffusion coefficient $2\pi T D_s$ for $D^0$ meson plotted as a function of $\rho_B/\rho_0$ for different values of temperatures and $I_a$ values where the dash-dotted lines represent the scenario based on the dilute gas assumption, modeled through relaxation time Eq. \eqref{eq_tau_c_a} and solid lines correspond to the degenerate Fermi liquid scenario, which is modeled through \eqref{eq:tau_b} along with the dashed line at $2\pi T D_s = 1$.}
	\label{fig:2piTDs}
\end{figure*}
\begin{figure*}
	\includegraphics[width=\linewidth]{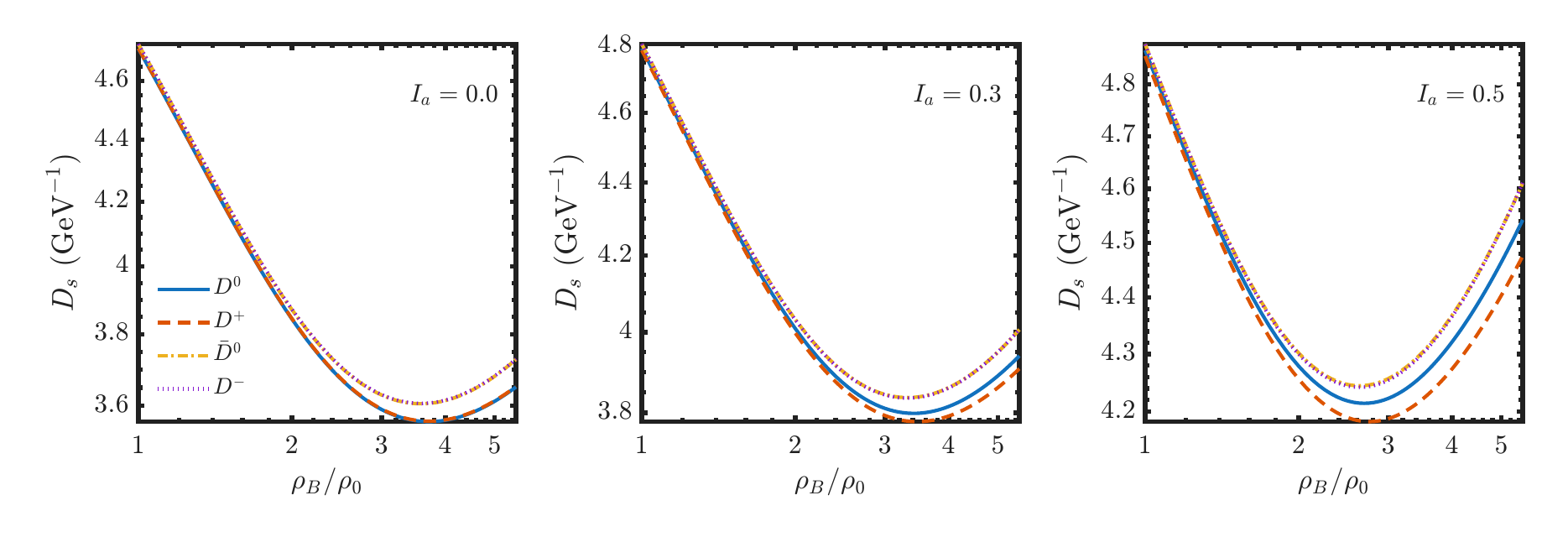}
	\caption{Spatial diffusion coefficient $D_s$ of the four $D$-meson species as a function of $\rho_B/\rho_0$ at $T = 100$ MeV for $I_a = 0$, $0.3$, and $0.5$ in the range $1-4\rho_0$ where the relaxation time is modeled through \eqref{eq:tau_b}.}
	\label{fig:Ds_species}
\end{figure*}

Before we present the results for diffusion coefficients in the dense medium, we compile previous estimations of the spatial diffusion coefficient of $D$ mesons through the hadronic medium at zero (baryon) density. Fig.~\ref{fig:data} shows a comparison of the scaled spatial diffusion coefficient $2\pi T D_s$ (where the thermal wavelength $2\pi T$ is multiplied to the spatial diffusion coefficient to make it dimensionless) of $D$ mesons in the hadronic medium produced in heavy-ion collisions, as obtained in various studies by He \textit{et al.}~\cite{He:2011yi}, Ghosh \textit{et al.}~\cite{Ghosh:2011bw}, Das \textit{et al.}~\cite{Das:2011vba}\footnote{In Refs. \cite{Satapathy:2022xdw,Dwibedi:2024amt}, the data taken from Ref.~\cite{Das:2011vba} was included but mistakenly cited as the Ref.~\cite{Ghosh:2011bw}.}, Ozvenchuk \textit{et al.}~\cite{Ozvenchuk:2014rpa}, Torres-Rincon \textit{et al.}~\cite{Torres-Rincon:2021yga}, Goswami \textit{et al.}~\cite{Goswami:2023hdl}, and Dwibedi \textit{et al.}~\cite{Dwibedi:2024amt}, plotted as a function of temperature $T$. A horizontal line at  $2\pi TD_s=1$ is also shown, marking the strong-coupling/holographic bound predicted by the AdS/CFT correspondence~\cite{Policastro:2002se}. The solid line corresponds to the diffusion coefficients of $D$ mesons through a nuclear medium computed with an average scattering length $a = 1.8$~fm. In contrast to the aforementioned studies, which consider a hadronic medium comprising a thermal gas of pions, kaons, and other light mesons, the medium in the present work is composed exclusively of protons and neutrons. The relaxation time $\tau_c$ is modeled following the approach of Dwibedi \textit{et al.}~\cite{Dwibedi:2024amt} through hard-sphere scattering, $\tau_c = 1/(n\,v\,\sigma_a)$, where $n$ is the total nucleon number density, $v$ is the average velocity of the $D$ meson, and $\sigma_a = \pi a^2$ is the scattering cross section with $a$ being the $D$--nucleon scattering length.

Although the spatial diffusion of the $D$ meson, along with the related momentum drag and diffusion coefficients, has been studied extensively, almost all such studies assume a baryon-free medium relevant to RHIC and LHC energies. However in a medium with non-zero baryochemical potential $\mu_B$ as those expected at the FAIR, the $D$ mesons are more thermalized with shorter relaxation times \cite{He:2011yi, Tolos:2013kva} and their transport properties have been explored previously in few studies including \cite{He:2011yi, Tolos:2013kva, Berrehrah:2014tva, Ozvenchuk:2014rpa,Goswami:2023hdl}. Here we outline a first systematic estimation of the spatial diffusion coefficient of $D$ meson through a dense nuclear medium from the kinetic theory approach. 

The relaxation rate of $D$ mesons in hadronic medium are often obtained from the elastic scattering amplitudes, effective field theoretical calculations or relaxation time approximations~\cite{He:2011yi,Ozvenchuk:2014rpa,Torres-Rincon:2021yga,Goswami:2023hdl,Dwibedi:2024amt}. In RTA the $D$ mesons are thought to move through a dilute gas of hadrons with collision rate given by $\tau_c^{-1} = n\,v\,\sigma_a$ as stated above \cite{Goswami:2023hdl,Dwibedi:2024amt}.  However, at low temperatures and high densities we expect this dilute gas to transition to a degenerate system approaching a Fermi liquid dictated by Pauli blocking and the exclusion principle. And to inspect this critically we simultaneously use two fits --- one corresponding to a dilute (ideal) gas and the other a degenerate Fermi liquid --- to model the nuclear medium and thus the relaxation time of the $D$ meson in the medium. 

In the first scenario, we consider a hard sphere scattering type of interaction between the particles of the medium, here, baryons -- protons and neutrons. We have the expression for the relaxation time $\tau_c$ as
\begin{equation}
	\tau_c^{(a)}=\frac{1}{n\sigma_a \langle v \rangle}\label{eq_tau_c_a}
\end{equation}
where $n$ and  $\sigma_a$ represent the total number density of the medium and the hard sphere scattering cross section of the $D$ meson. The total nucleon number density entering the scattering rate
includes particles and antiparticles for both proton and
neutron species,
\begin{equation}
	n=\displaystyle\sum_{N=p,\,n,\,\bar{p},\,\bar{n}}
	g_N\int \frac{d^3p}{(2\pi)^3}\;
	f_N	
	\label{eq:ntot}
\end{equation}
where $f_N = 1/[e^{(\sqrt{p^2 + m_N^{*2}}\mp\mu^*_N)/T} + 1]$ and $m_{N}^*$ and $\mu_{N}^*$ are the in-medium nucleon
masses and chemical potentials, which depend on the baryon density, temperature and
isospin asymmetry. The $D$ mesons, treated as bosonic quasi-particles with
vanishing chemical potential ($\mu_D = 0$), have an average velocity
\begin{equation}
	\langle v \rangle
	= \frac{g\displaystyle\int \frac{d^3p}
		{(2\pi)^3}\frac{p}{\sqrt{p^2+m_D^{*2}}}\,f_0}
	{g\displaystyle\int \frac{d^3p}
		{(2\pi)^3}\,f_0} \,,
	\label{eq:v_avg}
\end{equation}
where $f_0 = 1/(e^{\sqrt{p^2 + m_D^{*2}}/T} - 1)$ and $m_D^*$ is the in-medium mass of the $D$ meson. We have taken scattering length $a = 0.8$~fm. 

In the second parametrization we consider a model that encodes the Fermi-surface structure and the characteristic
$T^{-2}$ scaling of the quasi-particle scattering rate in a
degenerate system due to Pauli blocking \cite{ashcroft1976solid}. Considering the nuclear medium to be a degenerate Fermi gas, for each nucleon species,
\begin{equation}
	\tau_{c,N} = \frac{\mu_N^{*(0)}}{A\,T^2} \,,
	\qquad N = p,\,n \,,
	\label{eq:tau0}
\end{equation}
where $\mu_N^{*(0)}$ is the effective nucleon chemical potential or the Fermi energy evaluated at zero temperature, and the dimensionless proportionality constant $A$ is taken to be 12.\footnote{According to Ref.~\cite{ashcroft1976solid}, the dimensionless number $A$ can take any value within the range 1–100 for a degenerate Fermi gas system. We have chosen 12 as a value on the lower end that satisfies $2\pi TD_s\geq1$.}

The average relaxation time of the nucleons is obtained by weighting
with the Fermi--Dirac distribution of each nucleon species,
\begin{equation}
	\bar{\tau}_{c,N} = \frac{
		\displaystyle\sum_{N=p,\,n,\,\bar{p},\,\bar{n}}
		g_N\int \frac{d^3p}{(2\pi)^3}\;\tau_{c,N}\;
		f_{N}}{
		\displaystyle\sum_{N=p,\,n,\,\bar{p},\,\bar{n}}
		g_N\int \frac{d^3p}{(2\pi)^3}\;
		f_N	}.
	\label{tau_avg}
\end{equation}
The $D$-meson relaxation time can then be written as
\begin{equation}
	\tau_c^{(b)} = \frac{m_D^*}{T}\,\bar{\tau}_{c,N}  \,,
	\label{eq:tau_b}
\end{equation}
which incorporates the mass-to-temperature ratio reflecting the
kinematic suppression of heavy-particle relaxation~\cite{Moore:2004tg}. This $T^{-2}$ scaling of the relaxation time in the low temperature degenerate limit of relativistic nuclear matter is emphasized by Hakim and Mornas where the study was carried out using the Walecka model \cite{Hakim:1993zz,Mornas:1994cc}. Several similar estimates of collisions rates, relaxation times and transport coefficients in this regime can be found in literature including \cite{Danielewicz:1984kt,Muller:1980nmr,Kohler:1982wvu,Randrup:1979kt,Collins:1980fc,Kohler:1982vng}.

\begin{figure}
	\includegraphics[width=\linewidth]{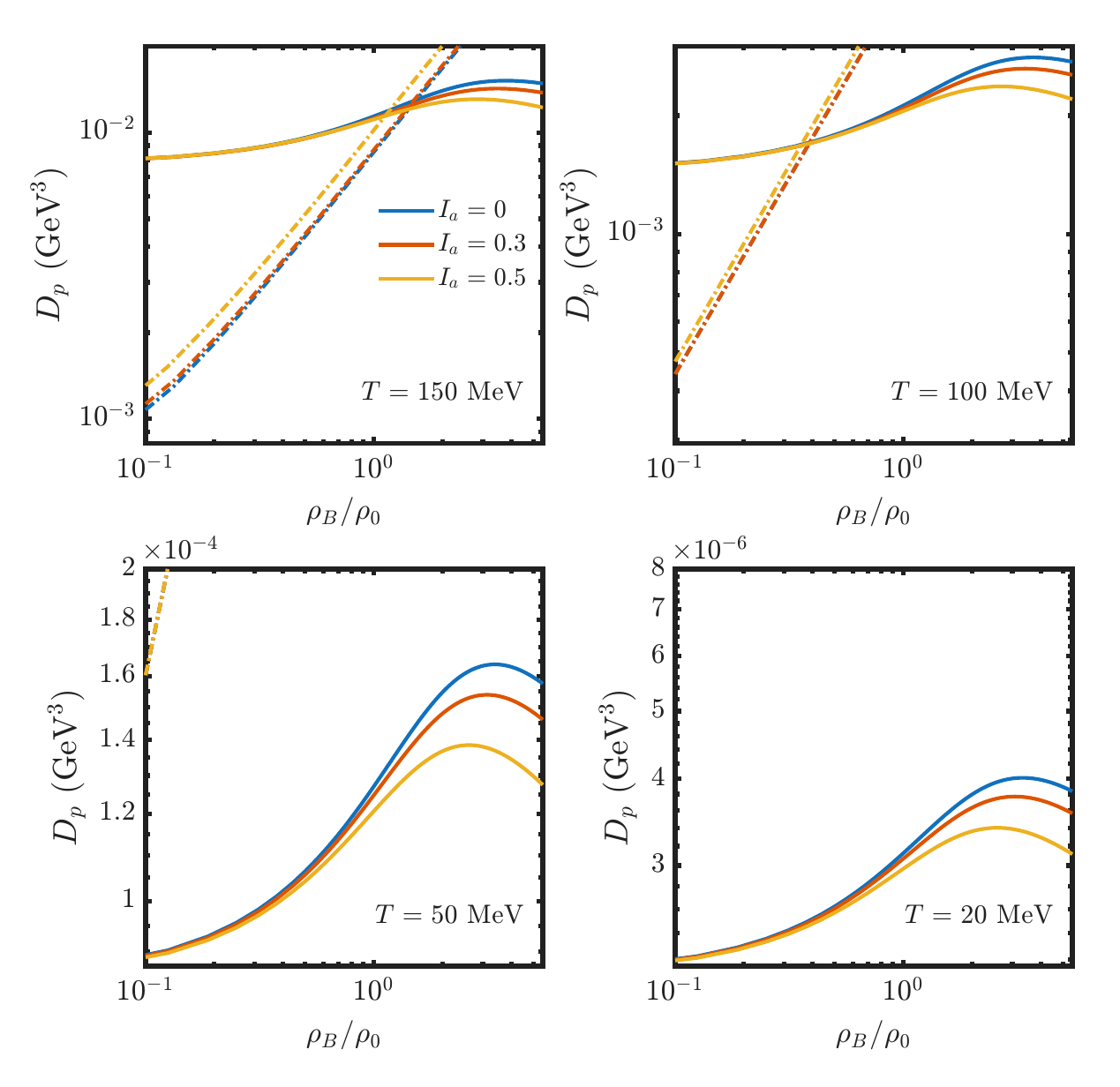}
	\caption{Momentum diffusion coefficient $D_p = T^2/D_s$ as a function of scaled baryon density $\rho_B/\rho_0$ for $D^0$ at different temperatures. The solid lines correspond to the degenerate Fermi liquid scenario, whereas the dash-dotted lines correspond to the dilute gas assumption.}
	\label{fig:D_mom}
\end{figure}

\begin{figure}
	\includegraphics[width=\linewidth]{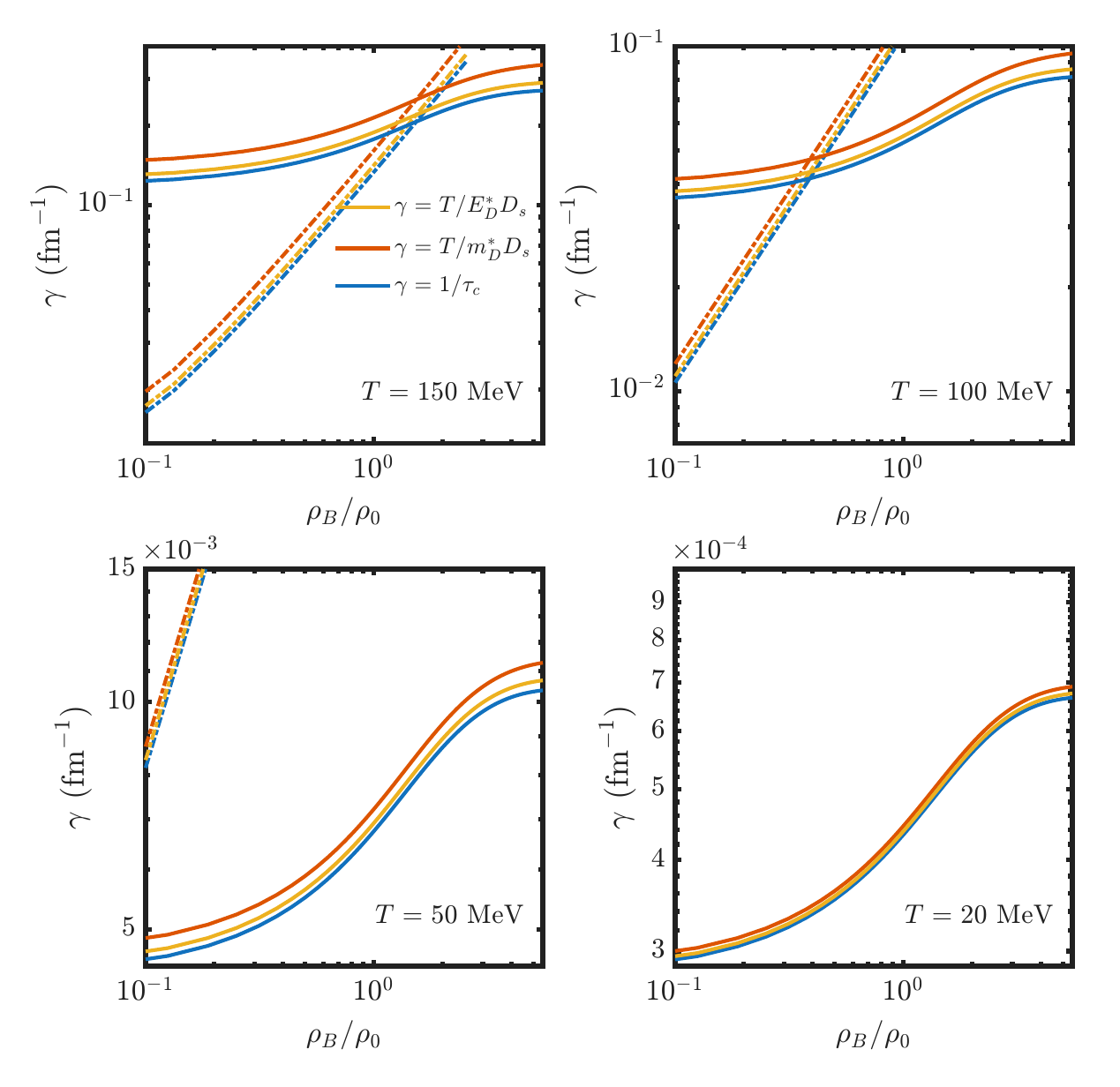}
	\caption{Momentum drag coefficient $\gamma$ as a function of scaled baryon density $\rho_B/\rho_0$ for $D^0$ at different temperatures. The solid lines correspond to the degenerate Fermi liquid scenario, whereas the dash-dotted lines correspond to the dilute gas assumption.}
	\label{fig:gamma}
\end{figure}

In the following discussion we calculate the transport coefficients of the $D$ meson using the above two parametrizations and compare the results to obtain estimates of diffusion coefficients as a function of baryon density. The spatial diffusion coefficient in the kinetic theory is given as the ratio of the $D$ meson conductivity to its susceptibility given by Eqs.~\eqref{sigma} and \eqref{chi}, respectively. We obtain the numerical results using the in-medium masses of the $D$ mesons and the relaxation time fits as detailed above. Fig.~\ref{fig:2piTDs} shows the scaled spatial diffusion coefficient $2\pi T D_s$ for $D^0$ meson plotted as a function of $\rho_B/\rho_0$ for different values of temperatures $T=20,50,100,150$ MeV with each panel corresponding to $I_a =0,0.3$ and $0.5$, respectively. Here, the dash-dotted lines are modeled using relaxation time $\tau_c^{(a)}$ given by Eq. \eqref{eq_tau_c_a} and solid lines are modeled through the relaxation time $\tau_c^{(b)}$ given by \eqref{eq:tau_b} along with the AdS/CFT bound $2\pi T D_s = 1$ (horizontal dashed line). We have provided extended plots of the dilute gas expression and the degenerate gas expression, which are respectively relevant in the low- and high-density domains. One can expect a transition from a dilute to a degenerate gas environment as one goes from the low- to the high-density domain. Our results, based on the chiral model, indicate that the rapidly decreasing trend of the spatial diffusion coefficient in the dilute gas domain transforms into a mild decreasing trend as it enters the degenerate gas domain. Within the degenerate gas domain, we also observe a smooth minimum or point of inflection in the scaled diffusion coefficient similar to the minimum obtained around the critical temperature in baryonless matter~\cite{Tolos:2013kva}. Upon increasing $I_a$, as shown in each panel of Fig.~\ref{fig:2piTDs}, we notice that the scaled spatial diffusion coefficient exhibit a slight but finite variation. Fig.~\ref{fig:Ds_species} shows a zoomed-in view of the minimum for different species and isospin asymmetry parameters at $T=100$ MeV. The splitting of diffusion coefficients among different species at high densities becomes increasingly pronounced with greater isospin asymmetry, consistent with their mass differences. The minimum in the curve tends to shift towards lower densities for medium with higher isospin asymmetry. The variation of momentum diffusion and drag coefficients can be obtained from the fluctuation dissipation relations relating the transport coefficients and they are depicted in Figs.~\ref{fig:D_mom} and~\ref{fig:gamma}. Here again, similar to Fig.~\ref{fig:2piTDs}, the solid lines illustrate the degenerate Fermi liquid scenario, whereas the dash-dotted lines depict the dilute gas assumption.

Fig.~\ref{fig:D_mom} depicts the momentum diffusion coefficient $D_p = T^2/D_s$ as a function of scaled baryon density $\rho_B/\rho_0$ for $D^0$ at different temperatures $T=150,100,50,20$ MeV for $I_a=0,0.3,0.5$. At $T = 150$ MeV, $D_p$ reaches values of order $10^{-2}$~GeV$^3$, while at $T = 20$ MeV it is suppressed by roughly three orders of magnitude.  The strong temperature dependence arises both from the explicit $T^2$ prefactor and from the implicit temperature dependence of $D_s$ through the Bose-Einstein distributions and in-medium masses. The effect of isospin asymmetry is visible as a splitting of the curves at each temperature -- larger $I_a$ leads to a slightly larger $D_p$, consistent with the reduced $D_s$ in neutron-rich matter. 

Fig.~\ref{fig:gamma} shows the momentum drag coefficient $\gamma$ as a function of scaled baryon density $\rho_B/\rho_0$ for $D^0$ at different temperatures $T=150,100,50,20$ MeV. We have simultaneously used three different expressions to obtain the drag coefficient. First, we obtain the drag coefficient assuming the fluctuation-dissipation relation, expressed as $\gamma=T/E_D^*D_s$ holds. We also use the definition of drag as the inverse of the relaxation time $\gamma=1/\tau_c$ and observe that both these values are close together but not coinciding. This can be attributed to the acausal behavior of the Navier-Stokes limit of hydrodynamics and the nature of relaxation time $\tau_c$ as a free parameter. We have also shown the values corresponding to $\gamma=T/m^*_DD_s$ -- the fluctuation-dissipation relation if the $D$ meson was truely non-relativistic with $p\rightarrow0$. 
Being inverses of the spatial diffusion coefficient $D_s$, the momentum drag $\gamma$ and diffusion $D_p$ coefficients show a rapid increasing trend in the low-density dilute gas domain and a mild increasing trend in the high-density degenerate gas domain. Near the (high) density range, where the spatial diffusion coefficient or the heavy meson relaxation time shows a mild valley structure, the momentum drag and diffusion coefficients exhibit a mild peak structure.

Overall, the effective QCD phase space derived from the chiral model can be linked to this non-trivial pattern of $D$ meson diffusion in the compressed baryonic matter (CBM) domain. This pattern depends on several parameters — temperature, density, isospin asymmetry, and the dilute-to-degenerate gas transition — which, to the best of our knowledge, are explored here for the first time. Given the absence of first-principle lattice QCD (LQCD) results in the dense sector, this effective-QCD-model-based estimate offers an important alternative theoretical insight, one that will be valuable for interpreting the corresponding experimental data expected from future facilities — FAIR at GSI, Germany, and NICA at JINR, Russia.

\section{Summary and conclusions}\label{summary}

We compute the transport properties of $D$ mesons in dense asymmetric nuclear matter using two complementary approaches for the collision time within kinetic theory. The relaxation time approximation (RTA) of the Boltzmann transport equation is employed to obtain the $D$ meson conductivity in the medium. The spatial diffusion coefficient is then expressed as the ratio of $D$ meson conductivity and susceptibility. For the main objective of this study is to observe the medium modifications on the $D$ meson as it traverses through a dense medium, we assume the background medium to be nucleonic -- consisting of protons and neutrons. The chiral SU(3) hadronic model is adopted and the medium modifications of the nucleons and the traversing $D$ mesons are studied in detail including their interactions. The in-medium $D$-meson masses decrease with increasing	baryon density, with an isospin splitting that grows with the asymmetry parameter $I_a$. The relaxation time of the $D$ meson dictates its diffusion through the medium and we critically examine this relationship here. A background of dilute hadron gas is shown to explain the diffusion of heavy mesons in a nearly baryonless medium with collision rate modeled as hard sphere scatterings. However, it is argued that this is no longer the case as the medium becomes dense and baryons dominate the scattering. As a benchmark of this behavior we consider a degenerate Fermi gas of nucleons scattering against each other and obtain the relaxation time within the ambit of Pauli blocking and exclusion principle. A $1/T^2$ suppression is expected in the relaxation time as the temperature decreases in this model. The Fermi energy and available phase space also influences the collision rate. The effective masses and chemical potential computed within the mean-field chiral model results in non-trivial behavior of the relaxation time and the spatial diffusion of $D$ mesons in the background of dense nuclear matter. The spatial diffusion coefficient $D_s$ decreases with density and then saturates or increases slightly following the non-monotonic behavior of effective chemical potential. A strong suppression of available phase space with lowering temperature results in rapidly increasing spatial diffusion coefficient.

Identifying the low-to-high density transformation as the dilute-to-degenerate gas transition of nuclear matter, we notice that the relaxation time and spatial diffusion coefficient of the heavy meson follow a rapidly-to-mildly decreasing trend with density during this transition, with a mild minimum also observed at a high-density point within the degenerate gas domain. The inverse pattern is observed in the momentum diffusion and drag coefficients.
The effect of isospin asymmetry is noticeable at densities above the nuclear saturation density.
The entire temperature- and density-dependent profile of $D$ meson diffusion phenomena is governed by the chiral model-based QCD phase space. To the best of our knowledge, the present work is the first to address this dilute-to-degenerate gas transition profile of  $D$ meson diffusion along the density axis. The present study is important for the theoretical understanding of transport phenomena in the dense sector of QCD, where first-principle calculations like LQCD fail, and is also for explaining the corresponding future experimental data from compressed baryonic matter expected at future facilities like FAIR at GSI, Germany, and NICA at JINR, Russia.
A planned future extension also includes the evolution of heavy flavor, including charm quarks, through the dense medium, and the extraction of phenomenological signatures -- the nuclear suppression factor and the flow coefficients.

\section{Acknowledgements}

Authors thank Ashutosh Dwibedi, Anand Rai and Dhananjay Singh for helpful discussions. This work was supported in part by the Ministry of Education, Government of India (D.R.M., M.K.); the Board of Research in Nuclear Sciences (BRNS), Department of Atomic Energy (DAE), Government of India, under Grant No. 57/14/01/2024-BRNS/313 (S.G.); and the Anusandhan National Research Foundation (ANRF), Government of India, under the Science and Engineering Research Board–Core Research Grant (SERB-CRG) scheme (File No. CRG/2023/000557).

\bibliographystyle{unsrturl}
\bibliography{ref_}

\begin{thebibliography}{10}

\bibitem{Gossiaux:2009hr}
P.~B. Gossiaux and J.~Aichelin.
\newblock {Tomography of the Quark Gluon Plasma by Heavy Quarks}.
\newblock {\em J. Phys. G}, 36:064028, 2009.
\newblock \href {https://arxiv.org/abs/0901.2462} {\path{arXiv:0901.2462}},
  \href {https://doi.org/10.1088/0954-3899/36/6/064028}
  {\path{doi:10.1088/0954-3899/36/6/064028}}.

\bibitem{PhysRevC.82.014908}
Santosh~K. Das, Jan-e Alam, and Payal Mohanty.
\newblock Drag of heavy quarks in quark gluon plasma at energies available at
  the cern large hadron collider (lhc).
\newblock {\em Phys. Rev. C}, 82:014908, Jul 2010.
\newblock URL: \url{https://link.aps.org/doi/10.1103/PhysRevC.82.014908}, \href
  {https://doi.org/10.1103/PhysRevC.82.014908}
  {\path{doi:10.1103/PhysRevC.82.014908}}.

\bibitem{Rapp:2008qc}
Ralf Rapp and Hendrik van Hees.
\newblock {Heavy Quark Diffusion as a Probe of the Quark-Gluon Plasma}.
\newblock 3 2008.
\newblock \href {https://arxiv.org/abs/0803.0901} {\path{arXiv:0803.0901}}.

\bibitem{Prino:2016cni}
Francesco Prino and Ralf Rapp.
\newblock {Open Heavy Flavor in QCD Matter and in Nuclear Collisions}.
\newblock {\em J. Phys. G}, 43(9):093002, 2016.
\newblock \href {https://arxiv.org/abs/1603.00529} {\path{arXiv:1603.00529}},
  \href {https://doi.org/10.1088/0954-3899/43/9/093002}
  {\path{doi:10.1088/0954-3899/43/9/093002}}.

\bibitem{Dong:2019byy}
Xin Dong, Yen-Jie Lee, and Ralf Rapp.
\newblock {Open Heavy-Flavor Production in Heavy-Ion Collisions}.
\newblock {\em Ann. Rev. Nucl. Part. Sci.}, 69:417--445, 2019.
\newblock \href {https://arxiv.org/abs/1903.07709} {\path{arXiv:1903.07709}},
  \href {https://doi.org/10.1146/annurev-nucl-101918-023806}
  {\path{doi:10.1146/annurev-nucl-101918-023806}}.

\bibitem{Rapp:2018qla}
A.~Beraudo et~al.
\newblock {Extraction of Heavy-Flavor Transport Coefficients in QCD Matter}.
\newblock {\em Nucl. Phys. A}, 979:21--86, 2018.
\newblock \href {https://arxiv.org/abs/1803.03824} {\path{arXiv:1803.03824}},
  \href {https://doi.org/10.1016/j.nuclphysa.2018.09.002}
  {\path{doi:10.1016/j.nuclphysa.2018.09.002}}.

\bibitem{Tolos:2016slr}
Laura Tolos, Juan~M. Torres-Rincon, and Santosh~K. Das.
\newblock {Transport coefficients of heavy baryons}.
\newblock {\em Phys. Rev. D}, 94(3):034018, 2016.
\newblock \href {https://arxiv.org/abs/1601.03743} {\path{arXiv:1601.03743}},
  \href {https://doi.org/10.1103/PhysRevD.94.034018}
  {\path{doi:10.1103/PhysRevD.94.034018}}.

\bibitem{Das:2024vac}
Santosh~K. Das, Juan~M. Torres-Rincon, and Ralf Rapp.
\newblock {Charm and bottom hadrons in hot hadronic matter}.
\newblock {\em Phys. Rept.}, 1129-1131:1--53, 2025.
\newblock \href {https://arxiv.org/abs/2406.13286} {\path{arXiv:2406.13286}},
  \href {https://doi.org/10.1016/j.physrep.2025.05.002}
  {\path{doi:10.1016/j.physrep.2025.05.002}}.

\bibitem{Zhao:2020jqu}
Jiaxing Zhao, Kai Zhou, Shile Chen, and Pengfei Zhuang.
\newblock {Heavy flavors under extreme conditions in high energy nuclear
  collisions}.
\newblock {\em Prog. Part. Nucl. Phys.}, 114:103801, 2020.
\newblock \href {https://arxiv.org/abs/2005.08277} {\path{arXiv:2005.08277}},
  \href {https://doi.org/10.1016/j.ppnp.2020.103801}
  {\path{doi:10.1016/j.ppnp.2020.103801}}.

\bibitem{Svetitsky:1987gq}
B.~Svetitsky.
\newblock {Diffusion of charmed quarks in the quark-gluon plasma}.
\newblock {\em Phys. Rev. D}, 37:2484--2491, 1988.
\newblock \href {https://doi.org/10.1103/PhysRevD.37.2484}
  {\path{doi:10.1103/PhysRevD.37.2484}}.

\bibitem{Braaten:1991jj}
Eric Braaten and Markus~H. Thoma.
\newblock {Energy loss of a heavy fermion in a hot plasma}.
\newblock {\em Phys. Rev. D}, 44:1298--1310, 1991.
\newblock \href {https://doi.org/10.1103/PhysRevD.44.1298}
  {\path{doi:10.1103/PhysRevD.44.1298}}.

\bibitem{Braaten:1991we}
Eric Braaten and Markus~H. Thoma.
\newblock {Energy loss of a heavy quark in the quark - gluon plasma}.
\newblock {\em Phys. Rev. D}, 44(9):R2625, 1991.
\newblock \href {https://doi.org/10.1103/PhysRevD.44.R2625}
  {\path{doi:10.1103/PhysRevD.44.R2625}}.

\bibitem{Moore:2004tg}
Guy~D. Moore and Derek Teaney.
\newblock {How much do heavy quarks thermalize in a heavy ion collision?}
\newblock {\em Phys. Rev. C}, 71:064904, 2005.
\newblock \href {https://arxiv.org/abs/hep-ph/0412346}
  {\path{arXiv:hep-ph/0412346}}, \href
  {https://doi.org/10.1103/PhysRevC.71.064904}
  {\path{doi:10.1103/PhysRevC.71.064904}}.

\bibitem{Kaczmarek:1999mm}
Olaf Kaczmarek, Frithjof Karsch, Edwin Laermann, and Martin Lutgemeier.
\newblock {Heavy quark potentials in quenched QCD at high temperature}.
\newblock {\em Phys. Rev. D}, 62:034021, 2000.
\newblock \href {https://arxiv.org/abs/hep-lat/9908010}
  {\path{arXiv:hep-lat/9908010}}, \href
  {https://doi.org/10.1103/PhysRevD.62.034021}
  {\path{doi:10.1103/PhysRevD.62.034021}}.

\bibitem{Banerjee:2011ra}
Debasish Banerjee, Saumen Datta, Rajiv Gavai, and Pushan Majumdar.
\newblock {Heavy Quark Momentum Diffusion Coefficient from Lattice QCD}.
\newblock {\em Phys. Rev. D}, 85:014510, 2012.
\newblock \href {https://arxiv.org/abs/1109.5738} {\path{arXiv:1109.5738}},
  \href {https://doi.org/10.1103/PhysRevD.85.014510}
  {\path{doi:10.1103/PhysRevD.85.014510}}.

\bibitem{Bazavov:2014kva}
A.~Bazavov, Y.~Burnier, and P.~Petreczky.
\newblock {Lattice calculation of the heavy quark potential at non-zero
  temperature}.
\newblock {\em Nucl. Phys. A}, 932:117--121, 2014.
\newblock \href {https://arxiv.org/abs/1404.4267} {\path{arXiv:1404.4267}},
  \href {https://doi.org/10.1016/j.nuclphysa.2014.09.078}
  {\path{doi:10.1016/j.nuclphysa.2014.09.078}}.

\bibitem{Altenkort:2023oms}
Luis Altenkort, Olaf Kaczmarek, Rasmus Larsen, Swagato Mukherjee, Peter
  Petreczky, Hai-Tao Shu, and Simon Stendebach.
\newblock {Heavy Quark Diffusion from 2+1 Flavor Lattice QCD with 320~MeV Pion
  Mass}.
\newblock {\em Phys. Rev. Lett.}, 130(23):231902, 2023.
\newblock \href {https://arxiv.org/abs/2302.08501} {\path{arXiv:2302.08501}},
  \href {https://doi.org/10.1103/PhysRevLett.130.231902}
  {\path{doi:10.1103/PhysRevLett.130.231902}}.

\bibitem{Casalderrey-Solana:2006fio}
Jorge Casalderrey-Solana and Derek Teaney.
\newblock {Heavy quark diffusion in strongly coupled N=4 Yang-Mills}.
\newblock {\em Phys. Rev. D}, 74:085012, 2006.
\newblock \href {https://arxiv.org/abs/hep-ph/0605199}
  {\path{arXiv:hep-ph/0605199}}, \href
  {https://doi.org/10.1103/PhysRevD.74.085012}
  {\path{doi:10.1103/PhysRevD.74.085012}}.

\bibitem{Gubser:2006nz}
Steven~S. Gubser.
\newblock {Momentum fluctuations of heavy quarks in the gauge-string duality}.
\newblock {\em Nucl. Phys. B}, 790:175--199, 2008.
\newblock \href {https://arxiv.org/abs/hep-th/0612143}
  {\path{arXiv:hep-th/0612143}}, \href
  {https://doi.org/10.1016/j.nuclphysb.2007.09.017}
  {\path{doi:10.1016/j.nuclphysb.2007.09.017}}.

\bibitem{Mannarelli:2005pz}
M.~Mannarelli and R.~Rapp.
\newblock {Hadronic modes and quark properties in the quark-gluon plasma}.
\newblock {\em Phys. Rev. C}, 72:064905, 2005.
\newblock \href {https://arxiv.org/abs/hep-ph/0505080}
  {\path{arXiv:hep-ph/0505080}}, \href
  {https://doi.org/10.1103/PhysRevC.72.064905}
  {\path{doi:10.1103/PhysRevC.72.064905}}.

\bibitem{vanHees:2007me}
H.~van Hees, M.~Mannarelli, V.~Greco, and R.~Rapp.
\newblock {Nonperturbative heavy-quark diffusion in the quark-gluon plasma}.
\newblock {\em Phys. Rev. Lett.}, 100:192301, 2008.
\newblock \href {https://arxiv.org/abs/0709.2884} {\path{arXiv:0709.2884}},
  \href {https://doi.org/10.1103/PhysRevLett.100.192301}
  {\path{doi:10.1103/PhysRevLett.100.192301}}.

\bibitem{Riek:2010fk}
F.~Riek and R.~Rapp.
\newblock {Quarkonia and Heavy-Quark Relaxation Times in the Quark-Gluon
  Plasma}.
\newblock {\em Phys. Rev. C}, 82:035201, 2010.
\newblock \href {https://arxiv.org/abs/1005.0769} {\path{arXiv:1005.0769}},
  \href {https://doi.org/10.1103/PhysRevC.82.035201}
  {\path{doi:10.1103/PhysRevC.82.035201}}.

\bibitem{Liu:2017qah}
Shuai Y.~F. Liu and Ralf Rapp.
\newblock {$T$-matrix Approach to Quark-Gluon Plasma}.
\newblock {\em Phys. Rev. C}, 97(3):034918, 2018.
\newblock \href {https://arxiv.org/abs/1711.03282} {\path{arXiv:1711.03282}},
  \href {https://doi.org/10.1103/PhysRevC.97.034918}
  {\path{doi:10.1103/PhysRevC.97.034918}}.

\bibitem{He:2011yi}
Min He, Rainer~J. Fries, and Ralf Rapp.
\newblock {Thermal Relaxation of Charm in Hadronic Matter}.
\newblock {\em Phys. Lett. B}, 701:445--450, 2011.
\newblock \href {https://arxiv.org/abs/1103.6279} {\path{arXiv:1103.6279}},
  \href {https://doi.org/10.1016/j.physletb.2011.06.019}
  {\path{doi:10.1016/j.physletb.2011.06.019}}.

\bibitem{Ghosh:2011bw}
Sabyasachi Ghosh, Santosh~K Das, Sourav Sarkar, and Jan-e Alam.
\newblock {Dragging $D$ mesons by hot hadrons}.
\newblock {\em Phys. Rev. D}, 84:011503, 2011.
\newblock \href {https://arxiv.org/abs/1104.0163} {\path{arXiv:1104.0163}},
  \href {https://doi.org/10.1103/PhysRevD.84.011503}
  {\path{doi:10.1103/PhysRevD.84.011503}}.

\bibitem{Tolos:2013kva}
Laura Tolos and Juan~M. Torres-Rincon.
\newblock {D-meson propagation in hot dense matter}.
\newblock {\em Phys. Rev. D}, 88:074019, 2013.
\newblock \href {https://arxiv.org/abs/1306.5426} {\path{arXiv:1306.5426}},
  \href {https://doi.org/10.1103/PhysRevD.88.074019}
  {\path{doi:10.1103/PhysRevD.88.074019}}.

\bibitem{Ozvenchuk:2014rpa}
Vitalii Ozvenchuk, Juan~M. Torres-Rincon, Pol~B. Gossiaux, Laura Tolos, and
  Joerg Aichelin.
\newblock {$D$-meson propagation in hadronic matter and consequences for
  heavy-flavor observables in ultrarelativistic heavy-ion collisions}.
\newblock {\em Phys. Rev. C}, 90:054909, 2014.
\newblock \href {https://arxiv.org/abs/1408.4938} {\path{arXiv:1408.4938}},
  \href {https://doi.org/10.1103/PhysRevC.90.054909}
  {\path{doi:10.1103/PhysRevC.90.054909}}.

\bibitem{Berrehrah:2014kba}
Hamza Berrehrah, Pol-Bernard Gossiaux, J{\"o}rg Aichelin, Wolfgang Cassing, and
  Elena Bratkovskaya.
\newblock {Dynamical collisional energy loss and transport properties of on-
  and off-shell heavy quarks in vacuum and in the Quark Gluon Plasma}.
\newblock {\em Phys. Rev. C}, 90(6):064906, 2014.
\newblock \href {https://arxiv.org/abs/1405.3243} {\path{arXiv:1405.3243}},
  \href {https://doi.org/10.1103/PhysRevC.90.064906}
  {\path{doi:10.1103/PhysRevC.90.064906}}.

\bibitem{Berrehrah:2014tva}
H.~Berrehrah, P.~B. Gossiaux, J.~Aichelin, W.~Cassing, J.~M. Torres-Rincon, and
  E.~Bratkovskaya.
\newblock {Transport coefficients of heavy quarks around $T_c$ at finite quark
  chemical potential}.
\newblock {\em Phys. Rev. C}, 90:051901, 2014.
\newblock \href {https://arxiv.org/abs/1406.5322} {\path{arXiv:1406.5322}},
  \href {https://doi.org/10.1103/PhysRevC.90.051901}
  {\path{doi:10.1103/PhysRevC.90.051901}}.

\bibitem{Scardina:2017ipo}
Francesco Scardina, Santosh~K. Das, Vincenzo Minissale, Salvatore Plumari, and
  Vincenzo Greco.
\newblock {Estimating the charm quark diffusion coefficient and thermalization
  time from D meson spectra at energies available at the BNL Relativistic Heavy
  Ion Collider and the CERN Large Hadron Collider}.
\newblock {\em Phys. Rev. C}, 96(4):044905, 2017.
\newblock \href {https://arxiv.org/abs/1707.05452} {\path{arXiv:1707.05452}},
  \href {https://doi.org/10.1103/PhysRevC.96.044905}
  {\path{doi:10.1103/PhysRevC.96.044905}}.

\bibitem{Torres-Rincon:2021yga}
Juan~M. Torres-Rincon, Gl{\`o}ria Monta{\~n}a, {\`A}ngels Ramos, and Laura
  Tolos.
\newblock {In-medium kinetic theory of D mesons and heavy-flavor transport
  coefficients}.
\newblock {\em Phys. Rev. C}, 105(2):025203, 2022.
\newblock \href {https://arxiv.org/abs/2106.01156} {\path{arXiv:2106.01156}},
  \href {https://doi.org/10.1103/PhysRevC.105.025203}
  {\path{doi:10.1103/PhysRevC.105.025203}}.

\bibitem{Goswami:2022szb}
Kangkan Goswami, Dushmanta Sahu, and Raghunath Sahoo.
\newblock {Understanding the QCD medium by the diffusion of charm quarks using
  a color string percolation model}.
\newblock {\em Phys. Rev. D}, 107(1):014003, 2023.
\newblock \href {https://arxiv.org/abs/2206.13786} {\path{arXiv:2206.13786}},
  \href {https://doi.org/10.1103/PhysRevD.107.014003}
  {\path{doi:10.1103/PhysRevD.107.014003}}.

\bibitem{Goswami:2023hdl}
Kangkan Goswami, Kshitish~Kumar Pradhan, Dushmanta Sahu, and Raghunath Sahoo.
\newblock {Diffusion and fluctuations of open charmed hadrons in an interacting
  hadronic medium}.
\newblock {\em Phys. Rev. D}, 108(7):074011, 2023.
\newblock \href {https://arxiv.org/abs/2307.04396} {\path{arXiv:2307.04396}},
  \href {https://doi.org/10.1103/PhysRevD.108.074011}
  {\path{doi:10.1103/PhysRevD.108.074011}}.

\bibitem{Pooja:2023gqt}
Pooja, Santosh~K. Das, Vincenzo Greco, and Marco Ruggieri.
\newblock {Thermalization and isotropization of heavy quarks in a non-Markovian
  medium in high-energy nuclear collisions}.
\newblock {\em Phys. Rev. D}, 108(5):054026, 2023.
\newblock \href {https://arxiv.org/abs/2306.13749} {\path{arXiv:2306.13749}},
  \href {https://doi.org/10.1103/PhysRevD.108.054026}
  {\path{doi:10.1103/PhysRevD.108.054026}}.

\bibitem{Satapathy:2022xdw}
Sarthak Satapathy, Sudipan De, Jayanta Dey, and Sabyasachi Ghosh.
\newblock {Spatial diffusion of heavy quarks in a background magnetic field}.
\newblock {\em Phys. Rev. C}, 109(2):024904, 2024.
\newblock \href {https://arxiv.org/abs/2212.08933} {\path{arXiv:2212.08933}},
  \href {https://doi.org/10.1103/PhysRevC.109.024904}
  {\path{doi:10.1103/PhysRevC.109.024904}}.

\bibitem{Dwibedi:2024amt}
Ashutosh Dwibedi, Nandita Padhan, Dani Rose~J. Marattukalam, Arghya Chatterjee,
  Sudipan De, and Sabyasachi Ghosh.
\newblock {Effect of coriolis force on diffusion of D meson}.
\newblock {\em J. Phys. G}, 52(9):095101, 2025.
\newblock \href {https://arxiv.org/abs/2411.09983} {\path{arXiv:2411.09983}},
  \href {https://doi.org/10.1088/1361-6471/adf983}
  {\path{doi:10.1088/1361-6471/adf983}}.

\bibitem{Dwibedi:2025fnz}
Ashutosh Dwibedi, Dani Rose~J. Marattukalam, Nandita Padhan, Dipannita Das,
  Arghya Chatterjee, Sudipan De, and Sabyasachi Ghosh.
\newblock {Anisotropic spatial diffusion of heavy quarks and D mesons in a
  rotating medium}.
\newblock {\em J. Subatomic Part. Cosmol.}, 4:100138, 2025.
\newblock \href {https://doi.org/10.1016/j.jspc.2025.100138}
  {\path{doi:10.1016/j.jspc.2025.100138}}.

\bibitem{Fukushima:2015wck}
Kenji Fukushima, Koichi Hattori, Ho-Ung Yee, and Yi~Yin.
\newblock {Heavy Quark Diffusion in Strong Magnetic Fields at Weak Coupling and
  Implications for Elliptic Flow}.
\newblock {\em Phys. Rev. D}, 93(7):074028, 2016.
\newblock \href {https://arxiv.org/abs/1512.03689} {\path{arXiv:1512.03689}},
  \href {https://doi.org/10.1103/PhysRevD.93.074028}
  {\path{doi:10.1103/PhysRevD.93.074028}}.

\bibitem{Friman:2011zz}
Bengt Friman, Claudia Hohne, Jorn Knoll, Stefan Leupold, Jorgen Randrup, Ralf
  Rapp, and Peter Senger, editors.
\newblock {\em {The CBM physics book: Compressed baryonic matter in laboratory
  experiments}}, volume 814.
\newblock 2011.
\newblock \href {https://doi.org/10.1007/978-3-642-13293-3}
  {\path{doi:10.1007/978-3-642-13293-3}}.

\bibitem{NA60:2022sze}
C.~Ahdida et~al.
\newblock {Letter of Intent: the NA60+ experiment}.
\newblock 12 2022.
\newblock \href {https://arxiv.org/abs/2212.14452} {\path{arXiv:2212.14452}}.

\bibitem{Arnaldi:2025ikz}
Roberta Arnaldi.
\newblock {Future facilities: The CERN SPS}.
\newblock {\em EPJ Web Conf.}, 339:01009, 2025.
\newblock \href {https://arxiv.org/abs/2505.10286} {\path{arXiv:2505.10286}},
  \href {https://doi.org/10.1051/epjconf/202533901009}
  {\path{doi:10.1051/epjconf/202533901009}}.

\bibitem{Bazavov:2017dus}
A.~Bazavov et~al.
\newblock {The QCD Equation of State to $\mathcal{O}(\mu_B^6)$ from Lattice
  QCD}.
\newblock {\em Phys. Rev. D}, 95(5):054504, 2017.
\newblock \href {https://arxiv.org/abs/1701.04325} {\path{arXiv:1701.04325}},
  \href {https://doi.org/10.1103/PhysRevD.95.054504}
  {\path{doi:10.1103/PhysRevD.95.054504}}.

\bibitem{Elfner:2022iae}
Hannah Elfner and Berndt M{\"u}ller.
\newblock {The exploration of hot and dense nuclear matter: introduction to
  relativistic heavy-ion physics}.
\newblock {\em J. Phys. G}, 50(10):103001, 2023.
\newblock \href {https://arxiv.org/abs/2210.12056} {\path{arXiv:2210.12056}},
  \href {https://doi.org/10.1088/1361-6471/ace824}
  {\path{doi:10.1088/1361-6471/ace824}}.

\bibitem{Hosaka:2016ypm}
Atsushi Hosaka, Tetsuo Hyodo, Kazutaka Sudoh, Yasuhiro Yamaguchi, and Shigehiro
  Yasui.
\newblock {Heavy Hadrons in Nuclear Matter}.
\newblock {\em Prog. Part. Nucl. Phys.}, 96:88--153, 2017.
\newblock \href {https://arxiv.org/abs/1606.08685} {\path{arXiv:1606.08685}},
  \href {https://doi.org/10.1016/j.ppnp.2017.04.003}
  {\path{doi:10.1016/j.ppnp.2017.04.003}}.

\bibitem{Montana:2023sft}
Gloria Montana, Angels Ramos, Laura Tolos, and Juan~M. Torres-Rincon.
\newblock {Recent progress on in-medium properties of heavy mesons from
  finite-temperature EFTs}.
\newblock {\em Front. in Phys.}, 11:1250939, 2023.
\newblock \href {https://arxiv.org/abs/2307.03640} {\path{arXiv:2307.03640}},
  \href {https://doi.org/10.3389/fphy.2023.1250939}
  {\path{doi:10.3389/fphy.2023.1250939}}.

\bibitem{Francis:2015daa}
A.~Francis, O.~Kaczmarek, M.~Laine, T.~Neuhaus, and H.~Ohno.
\newblock {Nonperturbative estimate of the heavy quark momentum diffusion
  coefficient}.
\newblock {\em Phys. Rev. D}, 92(11):116003, 2015.
\newblock \href {https://arxiv.org/abs/1508.04543} {\path{arXiv:1508.04543}},
  \href {https://doi.org/10.1103/PhysRevD.92.116003}
  {\path{doi:10.1103/PhysRevD.92.116003}}.

\bibitem{Schlichting:2019abc}
Soeren Schlichting and Derek Teaney.
\newblock {The First fm/c of Heavy-Ion Collisions}.
\newblock {\em Ann. Rev. Nucl. Part. Sci.}, 69:447--476, 2019.
\newblock \href {https://arxiv.org/abs/1908.02113} {\path{arXiv:1908.02113}},
  \href {https://doi.org/10.1146/annurev-nucl-101918-023825}
  {\path{doi:10.1146/annurev-nucl-101918-023825}}.

\bibitem{Capellino:2022nvf}
F.~Capellino, A.~Beraudo, A.~Dubla, S.~Floerchinger, S.~Masciocchi,
  J.~Pawlowski, and I.~Selyuzhenkov.
\newblock {Fluid-dynamic approach to heavy-quark diffusion in the quark-gluon
  plasma}.
\newblock {\em Phys. Rev. D}, 106(3):034021, 2022.
\newblock \href {https://arxiv.org/abs/2205.07692} {\path{arXiv:2205.07692}},
  \href {https://doi.org/10.1103/PhysRevD.106.034021}
  {\path{doi:10.1103/PhysRevD.106.034021}}.

\bibitem{Holt:2014hma}
Jeremy~W. Holt, Mannque Rho, and Wolfram Weise.
\newblock {Chiral symmetry and effective field theories for hadronic, nuclear
  and stellar matter}.
\newblock {\em Phys. Rept.}, 621:2--75, 2016.
\newblock \href {https://arxiv.org/abs/1411.6681} {\path{arXiv:1411.6681}},
  \href {https://doi.org/10.1016/j.physrep.2015.10.011}
  {\path{doi:10.1016/j.physrep.2015.10.011}}.

\bibitem{Papazoglou:1997uw}
P.~Papazoglou, S.~Schramm, J.~Schaffner-Bielich, Horst Stoecker, and
  W.~Greiner.
\newblock {Chiral Lagrangian for strange hadronic matter}.
\newblock {\em Phys. Rev. C}, 57:2576--2588, 1998.
\newblock \href {https://arxiv.org/abs/nucl-th/9706024}
  {\path{arXiv:nucl-th/9706024}}, \href
  {https://doi.org/10.1103/PhysRevC.57.2576}
  {\path{doi:10.1103/PhysRevC.57.2576}}.

\bibitem{Papazoglou:1998vr}
P.~Papazoglou, D.~Zschiesche, S.~Schramm, J.~Schaffner-Bielich, Horst Stoecker,
  and W.~Greiner.
\newblock {Nuclei in a chiral SU(3) model}.
\newblock {\em Phys. Rev. C}, 59:411--427, 1999.
\newblock \href {https://arxiv.org/abs/nucl-th/9806087}
  {\path{arXiv:nucl-th/9806087}}, \href
  {https://doi.org/10.1103/PhysRevC.59.411}
  {\path{doi:10.1103/PhysRevC.59.411}}.

\bibitem{Mishra:2003tr}
A.~Mishra, K.~Balazs, D.~Zschiesche, S.~Schramm, Horst Stoecker, and
  W.~Greiner.
\newblock {Effects of Dirac sea polarization on hadronic properties: A Chiral
  SU(3) approach}.
\newblock {\em Phys. Rev. C}, 69:024903, 2004.
\newblock \href {https://arxiv.org/abs/nucl-th/0308064}
  {\path{arXiv:nucl-th/0308064}}, \href
  {https://doi.org/10.1103/PhysRevC.69.024903}
  {\path{doi:10.1103/PhysRevC.69.024903}}.

\bibitem{Zschiesche:2003qq}
D.~Zschiesche, A.~Mishra, S.~Schramm, Horst Stoecker, and W.~Greiner.
\newblock {In-medium vector meson masses in a chiral SU(3) model}.
\newblock {\em Phys. Rev. C}, 70:045202, 2004.
\newblock \href {https://arxiv.org/abs/nucl-th/0302073}
  {\path{arXiv:nucl-th/0302073}}, \href
  {https://doi.org/10.1103/PhysRevC.70.045202}
  {\path{doi:10.1103/PhysRevC.70.045202}}.

\bibitem{Mishra:2006wy}
Amruta Mishra and Stefan Schramm.
\newblock {Isospin dependent kaon and antikaon optical potentials in dense
  hadronic matter}.
\newblock {\em Phys. Rev. C}, 74:064904, 2006.
\newblock \href {https://arxiv.org/abs/nucl-th/0607050}
  {\path{arXiv:nucl-th/0607050}}, \href
  {https://doi.org/10.1103/PhysRevC.74.064904}
  {\path{doi:10.1103/PhysRevC.74.064904}}.

\bibitem{Marattukalam:2024mef}
Dani Rose~J. Marattukalam, Ashutosh Dwibedi, Sourodeep De, and Sabyasachi
  Ghosh.
\newblock {Possibility of quantum Hall effect in dense quark matter
  environments: A chiral model approach}.
\newblock {\em Phys. Rev. D}, 112(5):054024, 2025.
\newblock \href {https://arxiv.org/abs/2410.22890} {\path{arXiv:2410.22890}},
  \href {https://doi.org/10.1103/ntpl-b8yl} {\path{doi:10.1103/ntpl-b8yl}}.

\bibitem{Marattukalam:2025lpi}
Dani Rose~J. Marattukalam, Ashutosh Dwibedi, Sourodeep De, and Sabyasachi
  Ghosh.
\newblock {Quantized conductivity in chiral effective model}.
\newblock {\em J. Subatomic Part. Cosmol.}, 4:100116, 2025.
\newblock \href {https://doi.org/10.1016/j.jspc.2025.100116}
  {\path{doi:10.1016/j.jspc.2025.100116}}.

\bibitem{Kumar:2025rxj}
Rajesh Kumar, Joaquin Grefa, Konstantin Maslov, Yuhan Wang, Arvind Kumar, Ralf
  Rapp, Claudia Ratti, and Veronica Dexheimer.
\newblock {Interacting mesons as degrees of freedom in a chiral model}.
\newblock {\em Phys. Rev. D}, 111(7):074029, 2025.
\newblock \href {https://arxiv.org/abs/2503.03057} {\path{arXiv:2503.03057}},
  \href {https://doi.org/10.1103/PhysRevD.111.074029}
  {\path{doi:10.1103/PhysRevD.111.074029}}.

\bibitem{Rai:2025lxw}
Anand Rai, Dani Rose~J. Marattukalam, Prasanta Murmu, Ashutosh Dwibedi, Rishabh
  Sharma, and Sabyasachi Ghosh.
\newblock {Towards compressed baryonic matter densities: thermodynamics and
  transport coefficients}.
\newblock 12 2025.
\newblock \href {https://arxiv.org/abs/2512.20282} {\path{arXiv:2512.20282}}.

\bibitem{Mishra:2003se}
A.~Mishra, E.~L. Bratkovskaya, J.~Schaffner-Bielich, S.~Schramm, and Horst
  Stoecker.
\newblock {Mass modification of D meson in hot hadronic matter}.
\newblock {\em Phys. Rev. C}, 69:015202, 2004.
\newblock \href {https://arxiv.org/abs/nucl-th/0308082}
  {\path{arXiv:nucl-th/0308082}}, \href
  {https://doi.org/10.1103/PhysRevC.69.015202}
  {\path{doi:10.1103/PhysRevC.69.015202}}.

\bibitem{Kumar:2010gb}
Arvind Kumar and Amruta Mishra.
\newblock {D mesons and charmonium states in asymmetric nuclear matter at
  finite temperatures}.
\newblock {\em Phys. Rev. C}, 81:065204, 2010.
\newblock \href {https://arxiv.org/abs/1005.5018} {\path{arXiv:1005.5018}},
  \href {https://doi.org/10.1103/PhysRevC.81.065204}
  {\path{doi:10.1103/PhysRevC.81.065204}}.

\bibitem{Blaschke:2011yv}
D.~Blaschke, P.~Costa, and Yu.~L. Kalinovsky.
\newblock {D mesons at finite temperature and density in the PNJL model}.
\newblock {\em Phys. Rev. D}, 85:034005, 2012.
\newblock \href {https://arxiv.org/abs/1107.2913} {\path{arXiv:1107.2913}},
  \href {https://doi.org/10.1103/PhysRevD.85.034005}
  {\path{doi:10.1103/PhysRevD.85.034005}}.

\bibitem{Hayashigaki:2000es}
Arata Hayashigaki.
\newblock {Mass modification of D meson at finite density in QCD sum rule}.
\newblock {\em Phys. Lett. B}, 487:96--103, 2000.
\newblock \href {https://arxiv.org/abs/nucl-th/0001051}
  {\path{arXiv:nucl-th/0001051}}, \href
  {https://doi.org/10.1016/S0370-2693(00)00760-7}
  {\path{doi:10.1016/S0370-2693(00)00760-7}}.

\bibitem{Wang:2015uya}
Zhi-Gang Wang.
\newblock {Analysis of heavy mesons in nuclear matter with a QCD sum rule
  approach}.
\newblock {\em Phys. Rev. C}, $\textbf{92}$:065205, 2015.
\newblock \href {https://doi.org/10.1103/PhysRevC.92.065205}
  {\path{doi:10.1103/PhysRevC.92.065205}}.

\bibitem{Kumar:2019axp}
Rajesh Kumar and Arvind Kumar.
\newblock {Analysis of pseudoscalar and scalar $D$ mesons and charmonium decay
  width in hot magnetized asymmetric nuclear matter}.
\newblock {\em Phys. Rev. C}, 101(1):015202, 2020.
\newblock \href {https://arxiv.org/abs/1908.09172} {\path{arXiv:1908.09172}},
  \href {https://doi.org/10.1103/PhysRevC.101.015202}
  {\path{doi:10.1103/PhysRevC.101.015202}}.

\bibitem{Tolos:2004yg}
L.~Tolos, J.~Schaffner-Bielich, and A.~Mishra.
\newblock {Properties of D-mesons in nuclear matter within a self-consistent
  coupled-channel approach}.
\newblock {\em Phys. Rev. C}, 70:025203, 2004.
\newblock \href {https://arxiv.org/abs/nucl-th/0404064}
  {\path{arXiv:nucl-th/0404064}}, \href
  {https://doi.org/10.1103/PhysRevC.70.025203}
  {\path{doi:10.1103/PhysRevC.70.025203}}.

\bibitem{Tolos:2005ft}
L.~Tolos, J.~Schaffner-Bielich, and H.~Stoecker.
\newblock {D-mesons: In-medium effects at FAIR}.
\newblock {\em Phys. Lett. B}, 635:85--92, 2006.
\newblock \href {https://arxiv.org/abs/nucl-th/0509054}
  {\path{arXiv:nucl-th/0509054}}, \href
  {https://doi.org/10.1016/j.physletb.2006.02.045}
  {\path{doi:10.1016/j.physletb.2006.02.045}}.

\bibitem{Mizutani:2006vq}
T.~Mizutani and A.~Ramos.
\newblock {D mesons in nuclear matter: A DN coupled-channel equations
  approach}.
\newblock {\em Phys. Rev. C}, 74:065201, 2006.
\newblock \href {https://arxiv.org/abs/hep-ph/0607257}
  {\path{arXiv:hep-ph/0607257}}, \href
  {https://doi.org/10.1103/PhysRevC.74.065201}
  {\path{doi:10.1103/PhysRevC.74.065201}}.

\bibitem{Tolos:2009nn}
L.~Tolos, C.~Garcia-Recio, and J.~Nieves.
\newblock {The Properties of D and D* mesons in the nuclear medium}.
\newblock {\em Phys. Rev. C}, 80:065202, 2009.
\newblock \href {https://arxiv.org/abs/0905.4859} {\path{arXiv:0905.4859}},
  \href {https://doi.org/10.1103/PhysRevC.80.065202}
  {\path{doi:10.1103/PhysRevC.80.065202}}.

\bibitem{Petreczky:2005nh}
Peter Petreczky and Derek Teaney.
\newblock {Heavy quark diffusion from the lattice}.
\newblock {\em Phys. Rev. D}, 73:014508, 2006.
\newblock \href {https://arxiv.org/abs/hep-ph/0507318}
  {\path{arXiv:hep-ph/0507318}}, \href
  {https://doi.org/10.1103/PhysRevD.73.014508}
  {\path{doi:10.1103/PhysRevD.73.014508}}.

\bibitem{Romatschke:2017ejr}
Paul Romatschke and Ulrike Romatschke.
\newblock {\em {Relativistic Fluid Dynamics In and Out of Equilibrium}}.
\newblock Cambridge Monographs on Mathematical Physics. Cambridge University
  Press, 5 2019.
\newblock \href {https://arxiv.org/abs/1712.05815} {\path{arXiv:1712.05815}},
  \href {https://doi.org/10.1017/9781108651998}
  {\path{doi:10.1017/9781108651998}}.

\bibitem{Riek:2010py}
Felix Riek and Ralf Rapp.
\newblock {Selfconsistent Evaluation of Charm and Charmonium in the Quark-Gluon
  Plasma}.
\newblock {\em New J. Phys.}, 13:045007, 2011.
\newblock \href {https://arxiv.org/abs/1012.0019} {\path{arXiv:1012.0019}},
  \href {https://doi.org/10.1088/1367-2630/13/4/045007}
  {\path{doi:10.1088/1367-2630/13/4/045007}}.

\bibitem{Weinberg:1968de}
Steven Weinberg.
\newblock {Nonlinear realizations of chiral symmetry}.
\newblock {\em Phys. Rev.}, 166:1568--1577, 1968.
\newblock \href {https://doi.org/10.1103/PhysRev.166.1568}
  {\path{doi:10.1103/PhysRev.166.1568}}.

\bibitem{Bardeen:1969ra}
William~A. Bardeen and B.~W. Lee.
\newblock {Some considerations on nonlinear realizations of chiral su(3) x
  su(3)}.
\newblock {\em Phys. Rev.}, 177:2389--2397, 1969.
\newblock \href {https://doi.org/10.1103/PhysRev.177.2389}
  {\path{doi:10.1103/PhysRev.177.2389}}.

\bibitem{Hayano:2008vn}
Ryugo~S. Hayano and Tetsuo Hatsuda.
\newblock {Hadron properties in the nuclear medium}.
\newblock {\em Rev. Mod. Phys.}, 82:2949, 2010.
\newblock \href {https://arxiv.org/abs/0812.1702} {\path{arXiv:0812.1702}},
  \href {https://doi.org/10.1103/RevModPhys.82.2949}
  {\path{doi:10.1103/RevModPhys.82.2949}}.

\bibitem{Cruz-Camacho:2024odu}
Nikolas Cruz-Camacho, Rajesh Kumar, Mateus Reinke~Pelicer, Jeff Peterson,
  T.~Andrew Manning, Roland Haas, Veronica Dexheimer, and Jaquelyn
  Noronha-Hostler.
\newblock {Phase stability in the three-dimensional open-source code for the
  chiral mean-field model}.
\newblock {\em Phys. Rev. D}, 111(9):094030, 2025.
\newblock \href {https://arxiv.org/abs/2409.06837} {\path{arXiv:2409.06837}},
  \href {https://doi.org/10.1103/PhysRevD.111.094030}
  {\path{doi:10.1103/PhysRevD.111.094030}}.

\bibitem{Kaur:2024cfm}
Manpreet Kaur and Arvind Kumar.
\newblock {Kaons and antikaons in isospin asymmetric dense resonance matter at
  finite temperature}.
\newblock {\em Phys. Rev. D}, 110(11):114054, 2024.
\newblock \href {https://arxiv.org/abs/2410.15685} {\path{arXiv:2410.15685}},
  \href {https://doi.org/10.1103/PhysRevD.110.114054}
  {\path{doi:10.1103/PhysRevD.110.114054}}.

\bibitem{Zschiesche:1999gf}
D.~Zschiesche, P.~Papazoglou, C.~W. Beckmann, S.~Schramm, J.~Schaffner-Bielich,
  Horst Stoecker, and W.~Greiner.
\newblock {Chiral model for dense, hot and strange hadronic matter}.
\newblock {\em Nucl. Phys. A}, 663:737--740, 2000.
\newblock \href {https://arxiv.org/abs/nucl-th/9908072}
  {\path{arXiv:nucl-th/9908072}}, \href
  {https://doi.org/10.1016/S0375-9474(99)00707-1}
  {\path{doi:10.1016/S0375-9474(99)00707-1}}.

\bibitem{Mishra:2008dj}
Amruta Mishra, Arvind Kumar, Sambuddha Sanyal, and Stefan Schramm.
\newblock {Kaon and antikaon optical potentials in isospin asymmetric hyperonic
  matter}.
\newblock {\em Eur. Phys. J. A}, 41:205--213, 2009.
\newblock \href {https://arxiv.org/abs/0808.1937} {\path{arXiv:0808.1937}},
  \href {https://doi.org/10.1140/epja/i2009-10777-6}
  {\path{doi:10.1140/epja/i2009-10777-6}}.

\bibitem{Kaur:2025kjk}
Manpreet Kaur and Arvind Kumar.
\newblock {{\ensuremath{\phi}} meson properties in dense resonance matter at
  finite temperature}.
\newblock {\em Phys. Rev. D}, 112(1):014030, 2025.
\newblock \href {https://arxiv.org/abs/2505.07065} {\path{arXiv:2505.07065}},
  \href {https://doi.org/10.1103/h2ll-5js4} {\path{doi:10.1103/h2ll-5js4}}.

\bibitem{Mishra:2008kg}
Amruta Mishra, Stefan Schramm, and W.~Greiner.
\newblock {Kaons and antikaons in asymmetric nuclear matter}.
\newblock {\em Phys. Rev. C}, 78:024901, 2008.
\newblock \href {https://arxiv.org/abs/0802.0363} {\path{arXiv:0802.0363}},
  \href {https://doi.org/10.1103/PhysRevC.78.024901}
  {\path{doi:10.1103/PhysRevC.78.024901}}.

\bibitem{Mishra:2004te}
A.~Mishra, E.~L. Bratkovskaya, J.~Schaffner-Bielich, S.~Schramm, and Horst
  Stoecker.
\newblock {Kaons and antikaons in hot and dense hadronic matter}.
\newblock {\em Phys. Rev. C}, 70:044904, 2004.
\newblock \href {https://arxiv.org/abs/nucl-th/0402062}
  {\path{arXiv:nucl-th/0402062}}, \href
  {https://doi.org/10.1103/PhysRevC.70.044904}
  {\path{doi:10.1103/PhysRevC.70.044904}}.

\bibitem{Barnes:1992ca}
Ted Barnes and E.~S. Swanson.
\newblock {Kaon - nucleon scattering amplitudes and Z* enhancements from quark
  Born diagrams}.
\newblock {\em Phys. Rev. C}, 49:1166--1184, 1994.
\newblock \href {https://arxiv.org/abs/nucl-th/9212008}
  {\path{arXiv:nucl-th/9212008}}, \href
  {https://doi.org/10.1103/PhysRevC.49.1166}
  {\path{doi:10.1103/PhysRevC.49.1166}}.

\bibitem{Roder:2003uz}
Dirk Roder, Jorg Ruppert, and Dirk~H. Rischke.
\newblock {Chiral symmetry restoration in linear sigma models with different
  numbers of quark flavors}.
\newblock {\em Phys. Rev. D}, 68:016003, 2003.
\newblock \href {https://arxiv.org/abs/nucl-th/0301085}
  {\path{arXiv:nucl-th/0301085}}, \href
  {https://doi.org/10.1103/PhysRevD.68.016003}
  {\path{doi:10.1103/PhysRevD.68.016003}}.

\bibitem{Mishra:2008cd}
Amruta Mishra and Arindam Mazumdar.
\newblock {D-mesons in asymmetric nuclear matter}.
\newblock {\em Phys. Rev. C}, 79:024908, 2009.
\newblock \href {https://arxiv.org/abs/0810.3067} {\path{arXiv:0810.3067}},
  \href {https://doi.org/10.1103/PhysRevC.79.024908}
  {\path{doi:10.1103/PhysRevC.79.024908}}.

\bibitem{Kumar:2009xc}
Arvind Kumar and Amruta Mishra.
\newblock {D mesons in asymmetric nuclear matter at finite temperatures}.
\newblock 12 2009.
\newblock \href {https://arxiv.org/abs/0912.2477} {\path{arXiv:0912.2477}}.

\bibitem{Das:2011vba}
Santosh~K. Das, Sabyasachi Ghosh, Sourav Sarkar, and Jan-e Alam.
\newblock {Drag and diffusion coefficients of $B$ mesons in hot hadronic
  matter}.
\newblock {\em Phys. Rev. D}, 85:074017, 2012.
\newblock \href {https://arxiv.org/abs/1109.3359} {\path{arXiv:1109.3359}},
  \href {https://doi.org/10.1103/PhysRevD.85.074017}
  {\path{doi:10.1103/PhysRevD.85.074017}}.

\bibitem{Policastro:2002se}
Giuseppe Policastro, Dam~T. Son, and Andrei~O. Starinets.
\newblock {From AdS / CFT correspondence to hydrodynamics}.
\newblock {\em JHEP}, 09:043, 2002.
\newblock \href {https://arxiv.org/abs/hep-th/0205052}
  {\path{arXiv:hep-th/0205052}}, \href
  {https://doi.org/10.1088/1126-6708/2002/09/043}
  {\path{doi:10.1088/1126-6708/2002/09/043}}.

\bibitem{ashcroft1976solid}
NW~Ashcroft.
\newblock Solid state physics.
\newblock {\em Thomson Learning}, 39, 1976.

\bibitem{Hakim:1993zz}
Remi Hakim and Lysiane Mornas.
\newblock {Collective effects on transport coefficients of relativistic nuclear
  matter}.
\newblock {\em Phys. Rev. C}, 47:2846--2860, 1993.
\newblock \href {https://doi.org/10.1103/PhysRevC.47.2846}
  {\path{doi:10.1103/PhysRevC.47.2846}}.

\bibitem{Mornas:1994cc}
L.~Mornas.
\newblock {Transport coefficients of relativistic nuclear and neutron matter
  with in-medium effects}.
\newblock {\em Nucl. Phys. A}, 573:554--586, 1994.
\newblock \href {https://doi.org/10.1016/0375-9474(94)90231-3}
  {\path{doi:10.1016/0375-9474(94)90231-3}}.

\bibitem{Danielewicz:1984kt}
P.~Danielewicz.
\newblock {Transport properties of excited nuclear matter and the shock wave
  profile}.
\newblock {\em Phys. Lett. B}, 146:168--175, 1984.
\newblock \href {https://doi.org/10.1016/0370-2693(84)91010-4}
  {\path{doi:10.1016/0370-2693(84)91010-4}}.

\bibitem{Muller:1980nmr}
K.~H. M{\"u}ller.
\newblock {A nuclear matter approach to the mean free path of a nucleon in a
  system of colliding ions}.
\newblock {\em Phys. Lett. B}, 93:247--249, 1980.
\newblock \href {https://doi.org/10.1016/0370-2693(80)90505-5}
  {\path{doi:10.1016/0370-2693(80)90505-5}}.

\bibitem{Kohler:1982wvu}
H.~S. K{\"o}hler.
\newblock {Effect of two-body collisions on heavy ion dynamics}.
\newblock {\em Nucl. Phys. A}, 378:181--188, 1982.
\newblock \href {https://doi.org/10.1016/0375-9474(82)90389-X}
  {\path{doi:10.1016/0375-9474(82)90389-X}}.

\bibitem{Randrup:1979kt}
J.~Randrup.
\newblock {Equilibration in nuclear matter}.
\newblock {\em Nucl. Phys. A}, 314:429--453, 1979.
\newblock \href {https://doi.org/10.1016/0375-9474(79)90606-7}
  {\path{doi:10.1016/0375-9474(79)90606-7}}.

\bibitem{Collins:1980fc}
M.~T. Collins and J.~J. Griffin.
\newblock {Mean Free Path of Nucleons in a Fermi Gas at Finite Temperature}.
\newblock {\em Nucl. Phys. A}, 348:63--74, 1980.
\newblock \href {https://doi.org/10.1016/0375-9474(80)90545-X}
  {\path{doi:10.1016/0375-9474(80)90545-X}}.

\bibitem{Kohler:1982vng}
H.~S. K{\"o}hler.
\newblock {Hot spots in nuclei by microscopic transport theory}.
\newblock {\em Nucl. Phys. A}, 378:159--180, 1982.
\newblock \href {https://doi.org/10.1016/0375-9474(82)90388-8}
  {\path{doi:10.1016/0375-9474(82)90388-8}}.

\end{thebibliography}

\end{document}